\documentclass[12pt]{article}

\usepackage{graphics}
\usepackage{amsmath}
\usepackage{epsfig}
\usepackage{cite}

\usepackage[T2A]{fontenc} 
\usepackage{amssymb,bm,amsfonts,amsmath,textcomp,mathtext,indentfirst} 
\usepackage{misccorr,enumerate,cite,csquotes,tocloft,float,setspace,xcolor}
\usepackage{graphicx}
\usepackage[outdir=./]{epstopdf}
\graphicspath{{img/}}
\usepackage[singlelinecheck=false]{caption}
\title{Bottomonia production and polarization in the NRQCD with $k_T$-factorization. II: $\Upsilon(2S)$ and $\chi_b(2P)$ mesons}
\author{N.A.~Abdulov$^{1,\,2}$, A.V.~Lipatov$^{2,\,3}$}

\begin{document}

\maketitle

\begin{center}

{\it $^1$Faculty of Physics, Lomonosov Moscow State University, 119991 Moscow, Russia}\\
{\it $^2$Skobeltsyn Institute of Nuclear Physics, Lomonosov Moscow State University, 119991 Moscow, Russia}\\
{\it $^3$Joint Institute for Nuclear Research, 141980 Dubna, Moscow Region, Russia}

\end{center} 

\vspace{0.2cm}

\begin{center}

{\bf Abstract }

\end{center}

The $\Upsilon(2S)$ production and polarization at high energies is 
studied in the framework of $k_T$-factorization approach.
Our consideration is based on the non-relativistic 
QCD formalism for bound states formation and off-shell production
amplitudes for hard partonic 
subprocesses. The direct production mechanism, feed-down contributions 
from radiative $\chi_b(3P)$ and $\chi_b(2P)$ decays and contributions from 
$\Upsilon(3S)$ decays are taken into account.
The transverse momentum dependent gluon
densities in a proton were derived from the 
Ciafaloni-Catani-Fiorani-Marchesini
evolution equation as well as from the Kimber-Martin-Ryskin prescription. 
Treating the non-perturbative color octet transitions 
in terms of the mulitpole radiation theory, we extract 
the corresponding non-perturbative matrix elements  
for $\Upsilon(2S)$ and $\chi_b(2P)$ mesons
from a combined fit to $\Upsilon(2S)$ transverse momenta distributions 
measured by the CMS and ATLAS Collaborations 
at the LHC energies $\sqrt s = 7$ and $13$~TeV 
and from the relative production rate $R^{\chi_b(2P)}_{\Upsilon(2S)}$ 
measured by the LHCb Collaboration at $\sqrt s = 7$ and $8$~TeV. 
Then we apply the extracted values 
to investigate the polarization 
parameters $\lambda_\theta$, $\lambda_\phi$ and $\lambda_{\theta\phi}$,
which determine the $\Upsilon(2S)$ spin density matrix. 
Our predictions
have a good agreement with the currently 
available data within the theoretical and experimental uncertainties. 

\newpage

\section{Introduction} \indent

Since it was first observed, the production of charmonia and 
bottomonia in hadronic collisions remains a 
subject of considerable theoretical and experimental studies 
\cite{kramer,lansberg,bramilla,gongLi,maWang,butenschonkniehl,chaoMa,gongWan,maWang2,likhoded,likhoded2,zhangYu,hanMa,zhangSun,butenschonHe,biswal}.
The theoretical framework for the description 
of heavy quarkonia production and decays provided by the
non-relativistic QCD (NRQCD) factorization \cite{bodwinBraaten,choLeibovich}. 
This formalism implies a separation of perturbatively 
calculated short-distance cross-sections for
the production of $Q\bar Q$ pair in an intermediate 
Fock state ${}^{2S+1}L_J^{(a)}$ with spin $S$, 
orbital angular momentum $L$, total angular momentum $J$ and 
color representation $a$ from long-distance 
non-perturbative matrix elements (NMEs), which describe the 
transition of that intermediate $Q\bar Q$ state 
into a physical quarkonium via soft gluon radiation. 

However, NRQCD meets some difficulties in simultaneous 
description of  the charmonia and bottomonia production cross section and polarization data,
as we have explained in our previous paper \cite{upsilonI}.
So, for example, having the NMEs fixed from fitting the charmonia transverse momentum distributions,
one disagrees with the polarization observables:
if the dominant contribution comes from the gluon fragmentation into an octet $Q\bar Q$ pair, the outgoing meson 
must have strong transverse polarization. The latter disagrees with the latest data \cite{cmslam,cms3,lhcb3,cdf1,cdf2}, which show the unpolarized or 
even longitudinally polarized particles (so called ``the polarization puzzle'').
Moreover, the NMEs, obtained from the collider data, dramatically depend on the minimal transverse momentum 
used in the fits \cite{faccioli} and are incompatible with each other when obtained from fitting the different data sets. 

A potential solution to this problem was proposed \cite{baranov} 
in the framework of a model 
that interprets the soft final state gluon radiation 
as a series of color-electric dipole transitions.
In this way the NMEs are represented in an explicit form inspired
by the classical multipole radiation theory, that leads 
to unpolarized or only weakly polarized mesons either 
because of the cancellation 
between the ${}^3P_1^{(8)}$ and ${}^3P_2^{(8)}$ contributions
or as a result of two
successive color-electric $E1$ dipole transitions in the 
chain ${}^3S_1^{(8)} \rightarrow {}^3P_J^{(8)} \rightarrow {}^3S_1^{(1)}$.
This scenario was already successfully applied to describe 
the recent data 
on charmonia production and polarization \cite{baranovLip,baranovLip2}.
Of course, it is important to investigate the bottomonia 
production within the same framework.

The data on $\Upsilon(nS)$ and $\chi_b(mP)$ mesons have been reported recently by the 
CMS \cite{cms1,cms2}, ATLAS \cite{atlas} and LHCb \cite{lhcb1,lhcb2} Collaborations
at $\sqrt s = 7$, $8$ and $13$~TeV. As it was shown \cite{gongWang,wangMa,gongWan2,fengGong,hanMa2}, these data can be explained within the 
NRQCD, both in polarization and yield.
Our present study continues the line started in the previous paper \cite{upsilonI}
and here we consider the production and polarization of $\Upsilon(2S)$ mesons.
To preserve the consistency with our studies \cite{baranovLip,baranovLip2}, we apply the $k_T$-factorization QCD approach \cite{sumkt1,sumkt2}
to describe the perturbative production of the $b\bar b$ pair in the 
hard scattering subprocess.
This approach is based on the Balitsky-Fadin-Kuraev-Lipatov 
(BFKL) \cite{bfkl} or 
Ciafaloni-Catani-Fiorani-Marchesini (CCFM) \cite{ccfm} evolution 
equations, which resum 
large logarithmic terms proportional to $\ln s \sim \ln 1/x$, important 
at high energies 
(or, equivalently, at low longitudinal momentum fraction $x$ of proton 
carried by gluon).
Resummation of the terms $\alpha_s^n\ln^n1/x$, $\alpha_s^n\ln^n \mu^2/\Lambda_{\rm QCD}^2$ and 
$\alpha_s^n\ln^n1/x\ln^n \mu^2/\Lambda_{\rm QCD}^2$ up to all orders in the perturbative expansion
results in Transverse Momentum Dependent (TMD) gluon distributions, that generalize the 
factorization of hadronic amplitudes beyond the conventional (collinear) DGLAP-based approximation.
For the different aspects of the $k_T$-factorization approach the reader may consult the reviews \cite{angeles}.
We determine the NMEs for $\Upsilon(2S)$ and $\chi_b(2P)$ mesons 
from the $\Upsilon(2S)$ transverse momentum 
distributions measured by the CMS \cite{cms1,cms2} and ATLAS \cite{atlas} 
Collaborations at $\sqrt s = 7$ and $13$~TeV and from the
relative production ratio $R^{\chi_b(2P)}_{\Upsilon(2S)}$
measured by the LHCb Collaboration at $\sqrt s = 7$ and 
$8$~TeV \cite{lhcbr}. In the calculations we take into account  
 the feed-down contributions from $\chi_b(3P)$, $\chi_b(2P)$ and $\Upsilon(3S)$ decays.
Then, we make predictions for polarization
parameters $\lambda_\theta$, $\lambda_\phi$, $\lambda_{\theta\phi}$ 
(and frame-independent parameter $\tilde\lambda$), which determine 
the $\Upsilon(2S)$ spin density matrix and compare
them to the currently available data \cite{cmslam,cdf2}.

The outline of our paper is the following. In Section 2 
we briefly recall the basic steps of our calculations. 
In Section 3 we perform a numerical fit and extract the NMEs
from the LHC data. 
Then we test the compatibility of the extracted NMEs with the 
available LHCb data on $\Upsilon(2S)$ transverse momentum distributions and
Tevatron data on the $\Upsilon(2S)$ transverse momentum distributions and 
polarization. Our conclusions are collected in Section 4.

\section{Theoretical framework} \indent

In the present paper we follow mostly the same steps as in our previous paper \cite{upsilonI}.
Our consideration is based on the off-shell gluon-gluon fusion 
subprocesses that represents
the true leading order (LO) in QCD:
\begin{equation}
g^*(k_1) + g^*(k_2) \rightarrow \Upsilon[{}^3S_1^{(1)}](p) + g(k),
\end{equation}	
\begin{equation}
g^*(k_1) + g^*(k_2) \rightarrow \Upsilon[{}^1S_0^{(8)},{}^3S_1^{(8)},{}^3P_J^{(8)}](p),
\end{equation}	
\begin{equation}
g^*(k_1) + g^*(k_2) \rightarrow \chi_{bJ}(p)[{}^3P_J^{(1)},{}^3S_1^{(8)}] \rightarrow \Upsilon(p_1) + \gamma(p_2),
\end{equation}

\noindent where we listed all intermediate color states, 
$J = 0, 1$ or $2$ and the four-momenta of all particles 
are indicated in the parentheses.
The respective cross sections for $2 \rightarrow 2$ and $2 \rightarrow 1$ subprocesses are:
\begin{multline}
\sigma = \int{{1}\over{8\pi(x_1x_2s) F}}f_g(x_1,\bm{k}^2_{1T},\mu^2)f_g(x_2,\bm{k}^2_{2T},\mu^2)\times\\
\times \overline{|A(g^*+g^*\rightarrow {\cal Q} + g)|^2}d\bm{p}^2_{T}d\bm{k}^2_{1T}d\bm{k}^2_{2T}dydy_g{{d\phi_1}\over{2\pi}}{{d\phi_2}\over{2\pi}},
\end{multline}	
\begin{equation}
\sigma=\int{{{2\pi}\over{x_1x_2sF}}f_g(x_1,\bm{k}^2_{1T},\mu^2)f_g(x_2,\bm{k}^2_{2T},\mu^2)\overline{|A(g^*+g^*\rightarrow {\cal Q})|^2}}d\bm{k}^2_{1T}d\bm{k}^2_{2T}dy{{d\phi_1}\over{2\pi}}{{d\phi_2}\over{2\pi}},
\label{sigma}
\end{equation} 

\noindent where ${\cal Q}$ is the $\Upsilon$ and/or $\chi_b$ meson, 
$\phi_1$ and $\phi_2$ are the azimuthal angles of  
initial off-shell gluons having 
the longitudinal momentum fractions $x_1$ and $x_2$, $\bm{p}_{T}$ and $y$ are the transverse 
momentum and rapidity of produced mesons, $y_g$ is the rapidity of 
outgoing gluon and $F$ is the off-shell flux factor \cite{bycling}.
The initial off-shell gluons have non-zero transverse momenta $\bm{k}^2_{1T} \neq 0$, 
$\bm{k}^2_{2T} \neq 0$ and an admixture of longitudinal component in the polarization 
vectors. So, the gluon spin density matrix is taken in the form ${\sum {\epsilon^\mu\epsilon^{*\nu}}} = {\bm k}_T^\mu{\bm k}_T^\nu/{\bm k}_T^2$,
where ${\bm k}_T$ is the component of the gluon momentum perpendicular 
to the beam axis \cite{sumkt1,sumkt2}. In the collinear limit ${\bm k}_T \rightarrow 0$ this expression 
converges to the ordinary one $\sum{\epsilon^\mu\epsilon^{*\nu}} = - 1/2g^{\mu\nu}$. 
In all other respects, we follow the standard QCD Feynman rules. As usual, the hard production 
amplitudes contain spin and color projection operators \cite{colorcs} that guarantee the proper quantum 
numbers of the state under consideration (see, for example, \cite{baranovLip,baranovLip2,upsilonI} for more details). 

The formation of $b\bar b$ bound states need additional explanation.
We employ the mechanism\footnote{The mechanism \cite{baranov} is not connected to the choice of 
factorization scheme ($k_T$ or collinear), but represents a completely independent issue.} proposed in \cite{baranov} 
and used previously \cite{baranovLip,baranovLip2,upsilonI}.
A soft gluon with a small 
energy $E\sim\Lambda_{\rm QCD}$ is emitted after the hard 
interaction is over, bringing away the unwanted color and 
changing other quantum numbers of the 
produced CO system. 
In our calculations such soft gluon emission is described
by a classical multipole expansion, 
in which the electric dipole ($E1$) transition dominates\cite{batuninCho}. 
Only a single $E1$ transition is needed to transform a $P$-wave 
state into an $S$-wave state and the structure of the 
respective ${^3P_J^{(8)}}\to {^3S_1^{(1)}}+g$ amplitudes
is given by\cite{batuninCho}:
\begin{equation}
A({}^3P_0^{(8)} \rightarrow \Upsilon + g) \sim k_\mu^{(g)}p^{\rm (CO)\mu}\epsilon_\nu^{(\Upsilon)}\epsilon^{(g)\nu},
\end{equation}
\begin{equation}
A({}^3P_1^{(8)} \rightarrow \Upsilon + g) \sim e^{\mu\nu\alpha\beta}k_\mu^{(g)}\epsilon_\nu^{\rm (CO)}\epsilon_\alpha^{(\Upsilon)}\epsilon_\beta^{(g)},
\end{equation}
\begin{equation}
A({}^3P_2^{(8)} \rightarrow \Upsilon + g) \sim p_\mu^{\rm (CO)}\epsilon_{\alpha\beta}^{\rm (CO)}\epsilon_\alpha^{(\Upsilon)} \left[ k_\mu^{(g)}\epsilon_\beta^{(g)} - k_\beta^{(g)}\epsilon_\mu^{(g)} \right],
\end{equation}

\noindent where $p^{\rm (CO)}_\mu$, $k^{(g)}_\mu$, 
$\epsilon^{(\Upsilon)}_\mu$, $\epsilon^{(g)}_\mu$, 
$\epsilon^{\rm (CO)}_\mu$ and $\epsilon^{\rm (CO)}_\mu$ are the 
momenta and polarization vectors of corresponding particles 
and $e^{\mu\nu\alpha\beta}$ is the fully 
antisymmetric Levi-Civita tensor. 
The transformation of color-octet $S$-wave state into the
color-singlet $S$-wave state is treated as 
two successive $E1$ transitions ${}^3S_1^{(8)} \rightarrow {}^3P_J^{(8)} + g$, 
${}^3P_J^{(8)} \rightarrow {}^3S_1^{(1)} + g$ proceeding via either of 
three intermediate ${}^3P_J^{(8)}$ states with $J = 0,1,2$. 
For each of these transitions 
we apply the same expressions (6) --- (8).
The amplitudes (6) --- (8) lead to the fact that the final state 
bottomonia come unpolarized \cite{baranov}, either because of the cancellation between 
the ${}^3P_1^{(8)}$ and ${}^3P_2^{(8)}$ contributions or as a result of two successive 
$E1$ transitions. This property remains true irrespectively of the numerical values of NMEs 
and only follows from the spin algebra.
The expressions (6) --- (8) can be applied for both
gluons and photons (up to an overall color factor) and 
can be used to 
calculate the polarization variables in radiative decays
in feed-down processes.

As we did in our previous paper \cite{upsilonI}, we have tested several sets of TMD gluon densities in a proton. 
Two of them (A0 \cite{A0set} and JH'2013 set 1 \cite{jhs}) were obtained from CCFM equation 
where all input parameters were fitted to the proton structure function $F_2(x,Q^2)$. 
Besides that, we have tested a parametrization obtained within the Kimber-Martin-Ryskin (KMR) prescription \cite{KMR},
which provides a method to construct the TMD quark and gluon densities from conventional (collinear) distributions. 
For the input, we have used recent LO NNPDF3.1 set \cite{nnpdf}.       

The parton level calculations were performed using the Monte-Carlo event generator \textsc{pegasus} \cite{pegasus}. 	

\section{Numerical results} \indent

In the present paper we set the masses  
$m_{\Upsilon(2S)} = 10.02326$~GeV, $m_{\chi_{b1}(3P)} = 10.512$~GeV,
$m_{\chi_{b2}(3P)} = 10.522$~GeV, $m_{\chi_{b0}(2P)} = 10.232$~GeV, $m_{\chi_{b1}(2P)} = 10.255$~GeV, $m_{\chi_{b2}(2P)} = 10.268$~GeV~\cite{pdg} and adopt the 
usual non-relativistic approximation $m_b = m_{\cal Q}/2$
for the beauty quark mass, where $m_{\cal Q}$ is the mass of 
bottomonium $\cal Q$.
We set the branching ratios $B(\Upsilon(2S) \rightarrow \mu^+\mu^-) = 0.0193$, 
$B(\Upsilon(3S) \rightarrow \Upsilon(2S) + X) =  0.1060$,
$B(\chi_{b0}(2P)\rightarrow \Upsilon(2S) + \gamma) = 0.0138$,
$B(\chi_{b1}(2P)\rightarrow \Upsilon(2S) + \gamma) = 0.1810$,
$B(\chi_{b2}(2P) \rightarrow \Upsilon(2S) + \gamma) = 0.0890$ \cite{pdg},
$B(\chi_{b1}(3P)\rightarrow \Upsilon(2S) + \gamma) = 0.0368$ and
$B(\chi_{b2}(3P) \rightarrow \Upsilon(2S) + \gamma) = 0.0191$ \cite{hanMa2}. 
Note that there is no experimental data for branching 
ratios of $\chi_b(3P)$,
so the values above are the results of assumption\cite{hanMa2}
that the total decay widths of $\chi_b(mP)$ are approximately 
independent on $m$.
We use the one-loop formula for the QCD coupling $\alpha_s$ 
with $n_f = 4(5)$ quark flavours at $\Lambda_{\rm QCD} = 250(167)$~MeV 
for A0 (KMR) gluon density
and two-loop expression for $\alpha_s$ with $n_f = 4$ 
and $\Lambda_{\rm QCD} = 200$~MeV for JH'2013 set 1 one.  
These parameters were obtained from best description of the structure function $F_2(x, Q^2)$ \cite{A0set,jhs,KMR}.
We set color-singlet NMEs $\langle\mathcal{O}^{\Upsilon(2S)}[{}^{3}S_1^{(1)}]\rangle  = 4.15$ GeV$^3$ 
and $\langle\mathcal{O}^{\chi_{b0}(2P)}[{}^{3}P_0^{(1)}]\rangle  = 2.61$ GeV$^5$ 
as obtained from the potential model calculations \cite{eichten}.
All the NMEs for $\Upsilon(3S)$ and $\chi_b(3P)$ mesons were derived
in \cite{upsilonI}.

\subsection{Fit of color octet NMEs} \indent

We performed a global fit to the $\Upsilon(2S)$ production data at 
the LHC and determined the corresponding NMEs for 
both $\Upsilon(2S)$ and $\chi_b(2P)$ mesons.
We have included in the fitting procedure the $\Upsilon(2S)$ transverse 
momentum distributions measured by
the CMS \cite{cms1,cms2} and ATLAS \cite{atlas} Collaborations 
at $\sqrt s = 7$ and $13$~TeV. 
To determine NMEs for $\chi_b(2P)$ mesons, we also included into the fit 
the recent LHCb data \cite{lhcbr} on the radiative 
$\chi_b(2P) \to \Upsilon(2S) + \gamma$ decays 
collected at $\sqrt s = 7$ and $8$~TeV. 
We have excluded from our fit low $p_T$ region and consider 
only the data at
$p_T > p_T^{\rm cut} = 10$~GeV, where the NRQCD 
formalism is believed to be most reliable. 

\begin{figure}
\begin{center}
\includegraphics[width=8.0cm]{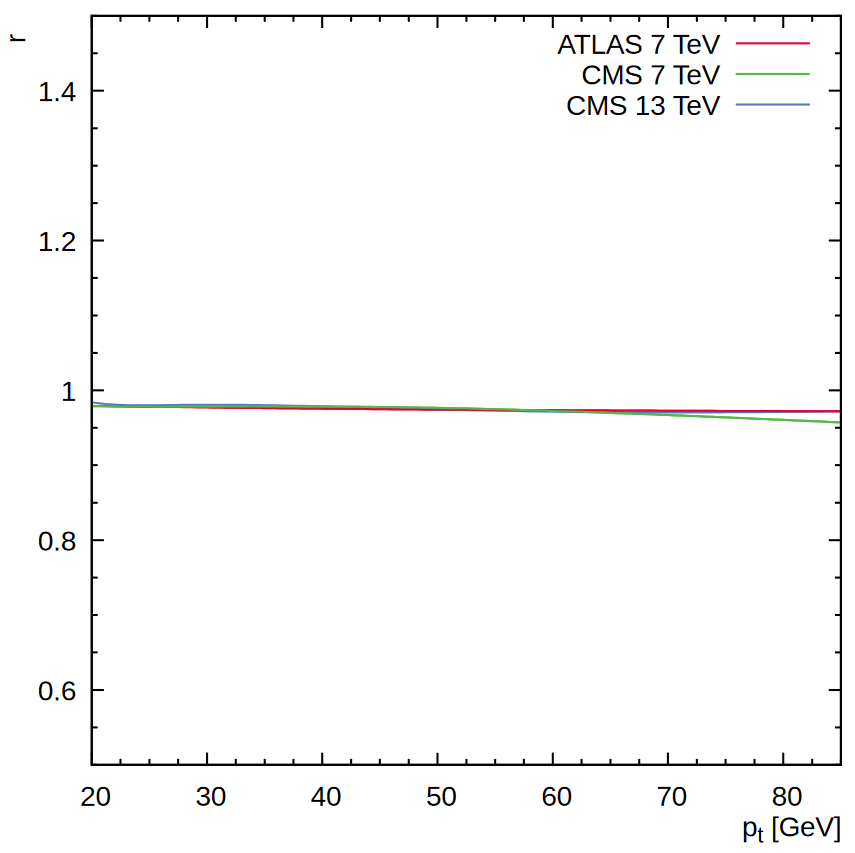}
\caption{The production ratio $r(p_t)$
calculated as a function of $\Upsilon(2S)$ transverse momentum $p_T$ in the 
different kinematical regions.}
\label{fig0}
\end{center}
\end{figure}

We would like to mention here a few important points.
First of all,
we found that the $p_T$ shape of the direct $\Upsilon[^3S_1^{(8)}]$ and 
feed-down $\chi_b[^3S_1^{(8)}]$ contributions to $\Upsilon(2S)$ production is almost 
the same in all kinematical regions
probed at the LHC. Thus, the ratio
\begin{equation}
r = { \sum\limits_{J = 0}^{2} (2J+1) \, B(\chi_{bJ}(2P) \to \Upsilon(2S) + \gamma) d\sigma[\chi_{bJ}(2P), {}^3S_1^{(8)}]/dp_T \over d\sigma [\Upsilon(2S), {}^3S_1^{(8)}]/dp_T }
\label{eqr}
\end{equation}

\noindent 
can be well approximated by a constant for a wide $\Upsilon(2S)$ 
transverse 
momentum $p_T$ and rapidity $y$ ranges at different energies, as seen in Fig.~\ref{fig0}. 
We estimate the mean-square average $r = 0.98 \pm 0.005$, which is 
practically independent on the TMD gluon density in a proton.
So that, we construct the linear combination
\begin{equation}
M_r = \langle\mathcal{O}^{\Upsilon(2S)}[{}^{3}S_1^{(8)}]\rangle + r \langle\mathcal{O}^{\chi_{b0}(2P)}[{}^{3}S_1^{(8)}]\rangle,
\end{equation}
\noindent
which can be only extracted from the measured $\Upsilon(2S)$ transverse
momentum distributions.
Then we use recent LHCb data \cite{lhcbr} on the ratio
of $\Upsilon(2S)$ mesons 
originating from the $\chi_b(2P)$ radiative decays measured
at $\sqrt s = 7$ and $8$~TeV:
\begin{equation}
R^{\chi_b(2P)}_{\Upsilon(2S)} = \sum\limits_{J = 0}^{2} {\sigma(pp\rightarrow \chi_{bJ}(2P) + X) \over \sigma(pp \rightarrow \Upsilon(2S) + X)} \times B(\chi_{bJ} \to \Upsilon(2S) + \gamma).
\label{eqrUp}
\end{equation}

\noindent
From the known $M_r$ and $R^{\chi_b(2P)}_{\Upsilon(2S)}$ values
one can separately determine the 
$\langle\mathcal{O}^{\Upsilon(2S)}[{}^{3}S_1^{(8)}]\rangle$ and 
$\langle\mathcal{O}^{\chi_{b0}(2P)}[{}^{3}S_1^{(8)}]\rangle$ and, therefore,
reconstruct full map of color octet NMEs for both $\Upsilon(2S)$ 
and $\chi_b(2P)$ mesons.

The fitting procedure was separately done in each of the rapidity 
subdivisions (using the fitting algorithm as implemented 
in the commonly used \textsc{gnuplot} package\cite{gnuplot}) 
under the requirement that all the NMEs be strictly positive. 
Then, the mean-square average of
the fitted values was taken. The corresponding uncertainties are 
estimated in the conventional way
using Student's t-distribution at the confidence level $P = 80$\%.
The results of our fits are collected in Table~1. For comparison, 
we also presented 
there the NMEs obtained in the conventional NLO NRQCD by other 
authors\cite{fengGong}.
The corresponding $\chi^2/d.o.f.$ 
are listed in Table~2, where we additionally
show their dependence on the minimal $\Upsilon(2S)$ transverse momenta 
involved into the fit $p_T^{\rm cut}$.
As one can see, the $\chi^2/d.o.f.$ tends to decrease when 
$p_T^{\rm cut}$ grows up and
best fit of the LHC data is achieved with A0 gluon, although 
other gluon densities also return reliable $\chi^2/d.o.f.$ values.
We note that including into the fit the latest CMS data \cite{cms2} taken at $\sqrt s = 13$ TeV leads to
2 --- 3 times higher values of $\chi^2/d.o.f.$ We have checked that this is true for both the $k_T$-factorization 
and collinear approaches\footnote{We have used the on-shell production amplitudes for color-octet $2 \to 2$ 
subprocesses from \cite{choLeibovich}.} and, therefore, it could be a sign of some inconsistency 
between these CMS data and all other measurements.
    
\begin{figure}
\begin{center}
\includegraphics[width=7.0cm]{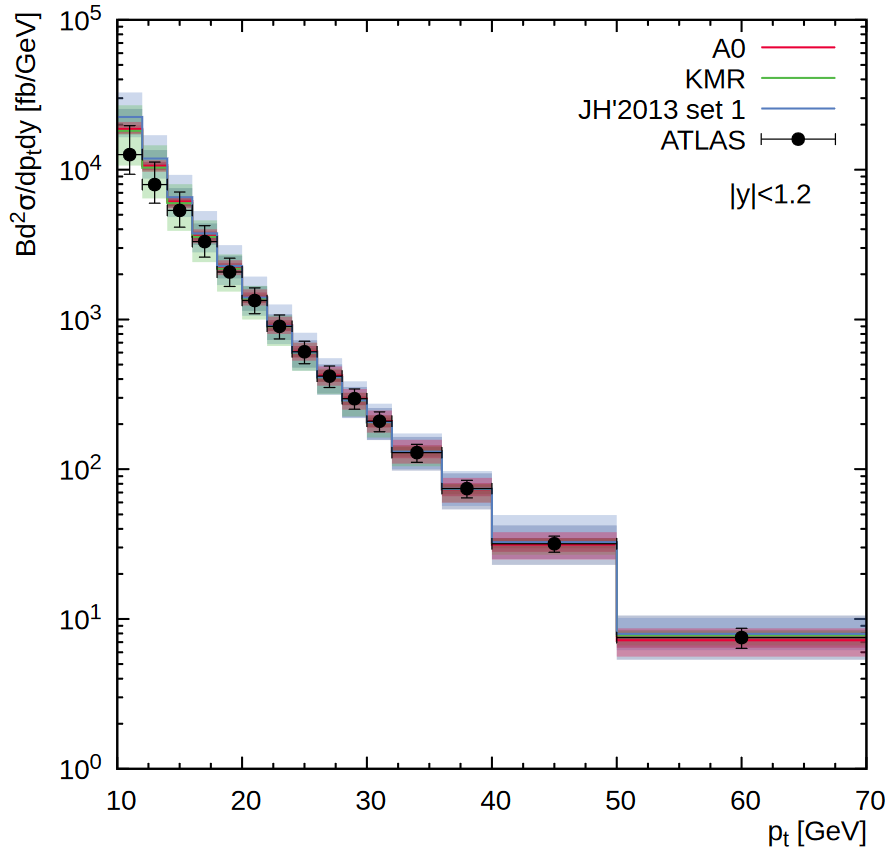}
\includegraphics[width=7.0cm]{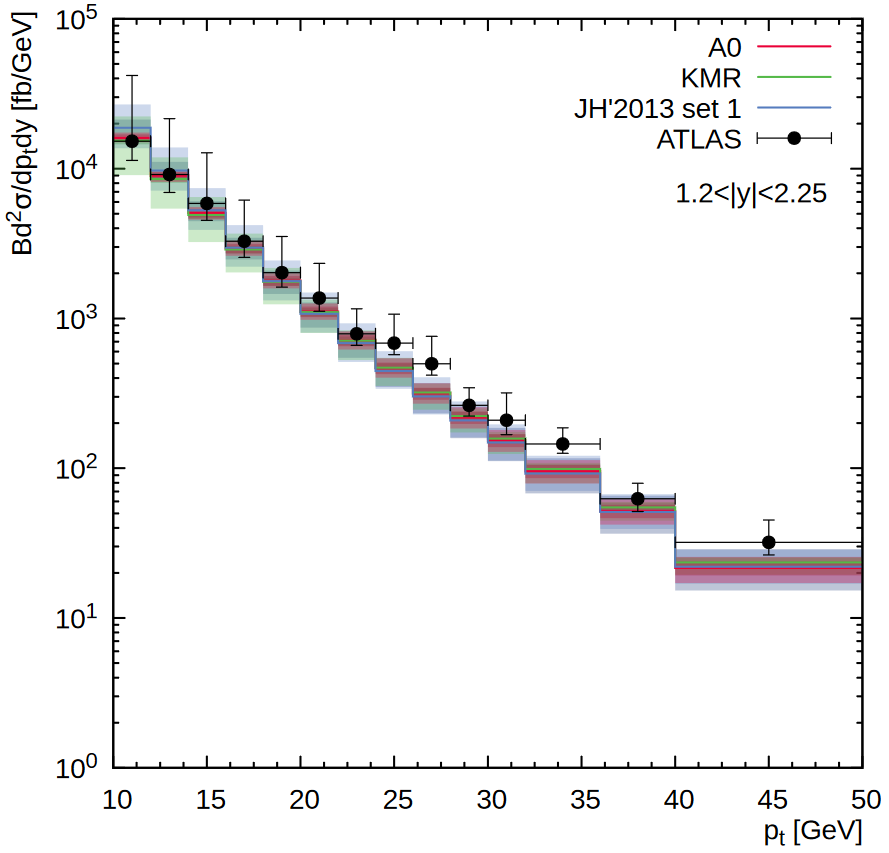}
\caption{Transverse momentum distribution of 
  inclusive $\Upsilon(2S)$ production calculated at 
  $\sqrt s = 7$~TeV in the different rapidity regions. 
  The red, green and blue histograms
  correspond to the predictions obtained with A0, KMR and JH'2013 set 1
  gluon densities. Shaded bands represent the total uncertainties 
  of our calculations, as it is described in text.
  The experimental data are from ATLAS \cite{atlas}.}
\label{fig1}
\end{center}
\end{figure}

\begin{figure}
\begin{center}
\includegraphics[width=7.0cm]{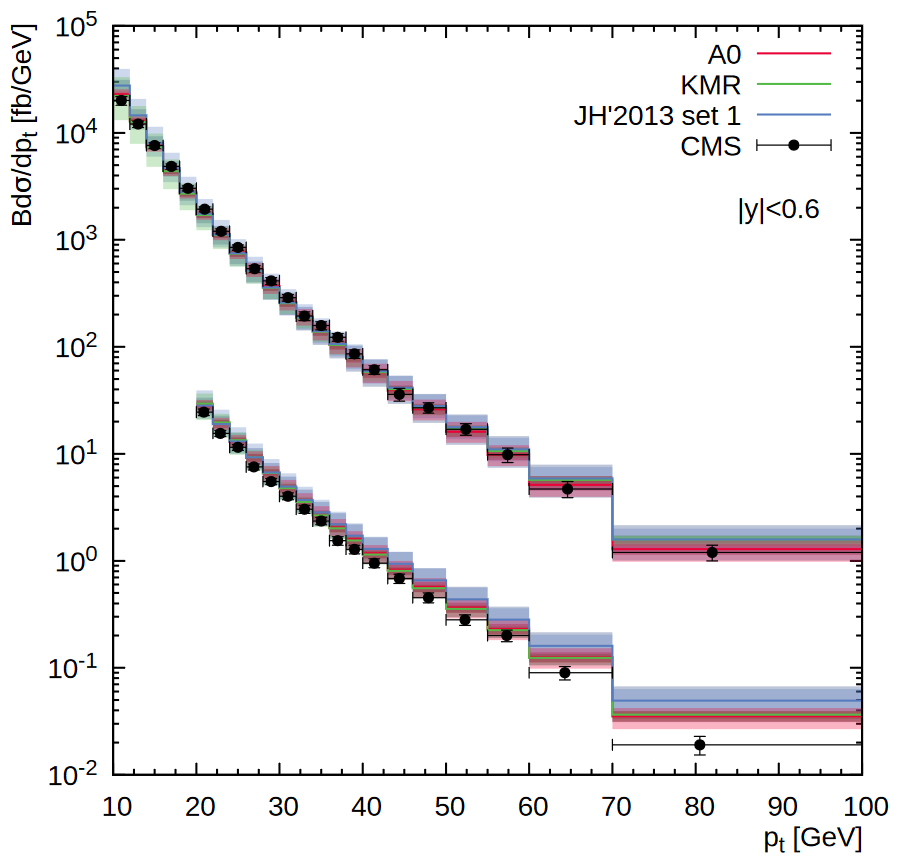}
\includegraphics[width=7.0cm]{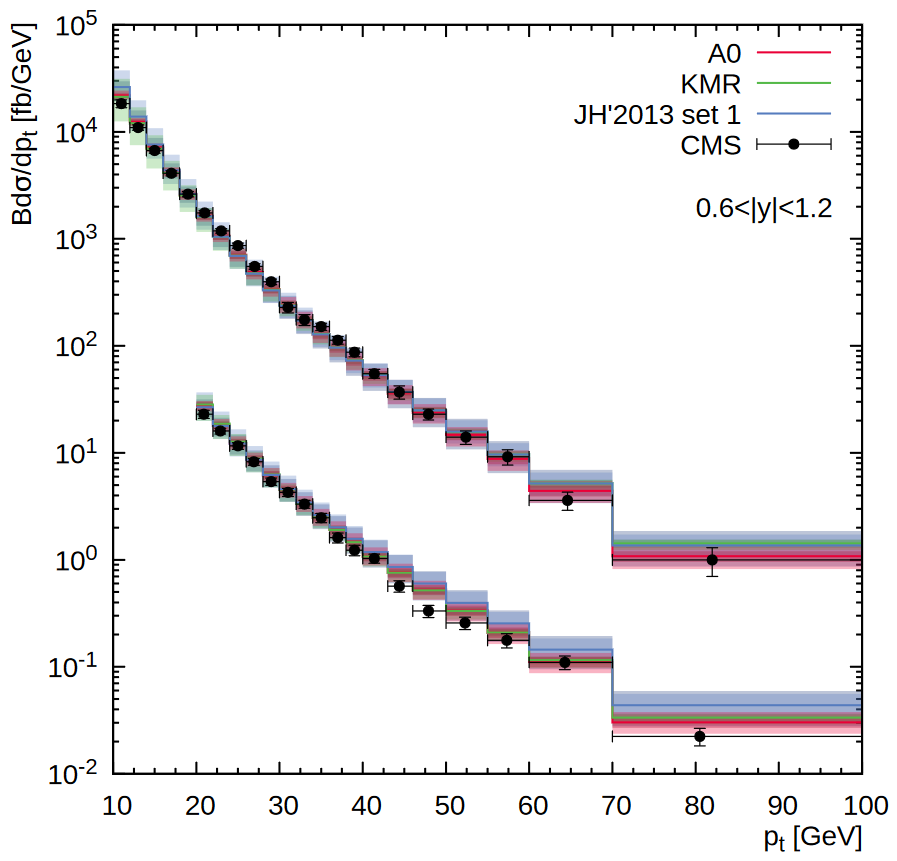}
\includegraphics[width=7.0cm]{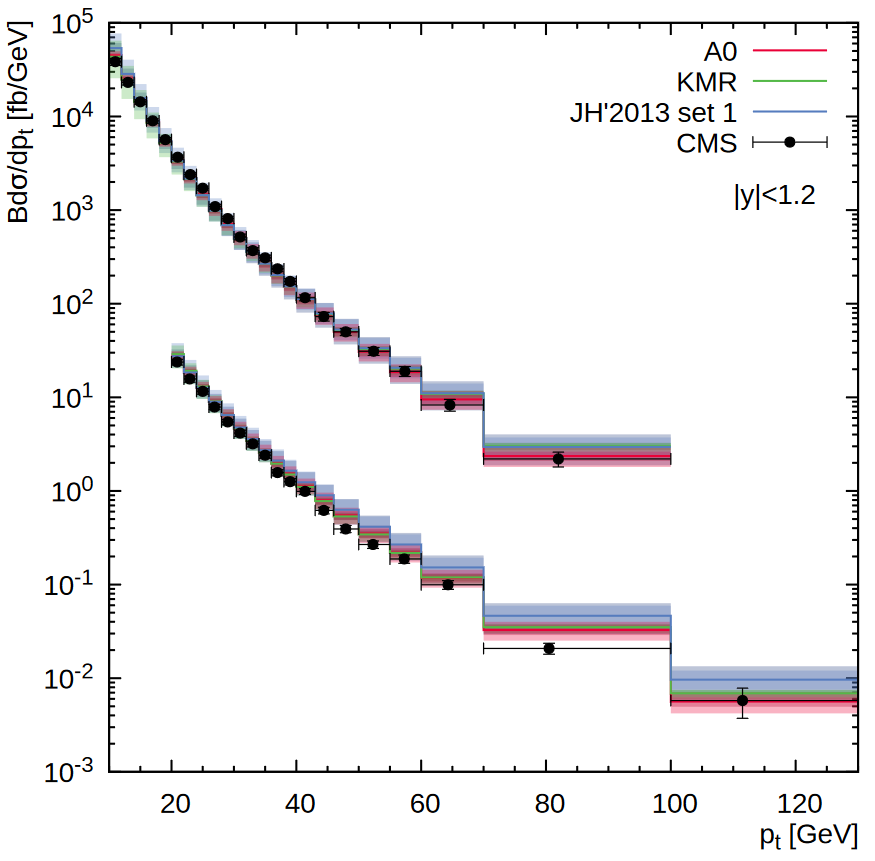}
\caption{Transverse momentum distribution of 
  inclusive $\Upsilon(2S)$ production calculated at $\sqrt s = 7$~TeV
  (upper histograms) and $\sqrt s = 13$~TeV (lower histograms, 
  divided by $100$) in the different rapidity regions. 
  Notation of all histograms is the same as in Fig.~\ref{fig1}.
  The experimental data are from CMS \cite{cms1,cms2}.}
\label{fig2}
\end{center}
\end{figure}

\begin{figure}
\begin{center}
\includegraphics[width=7.0cm]{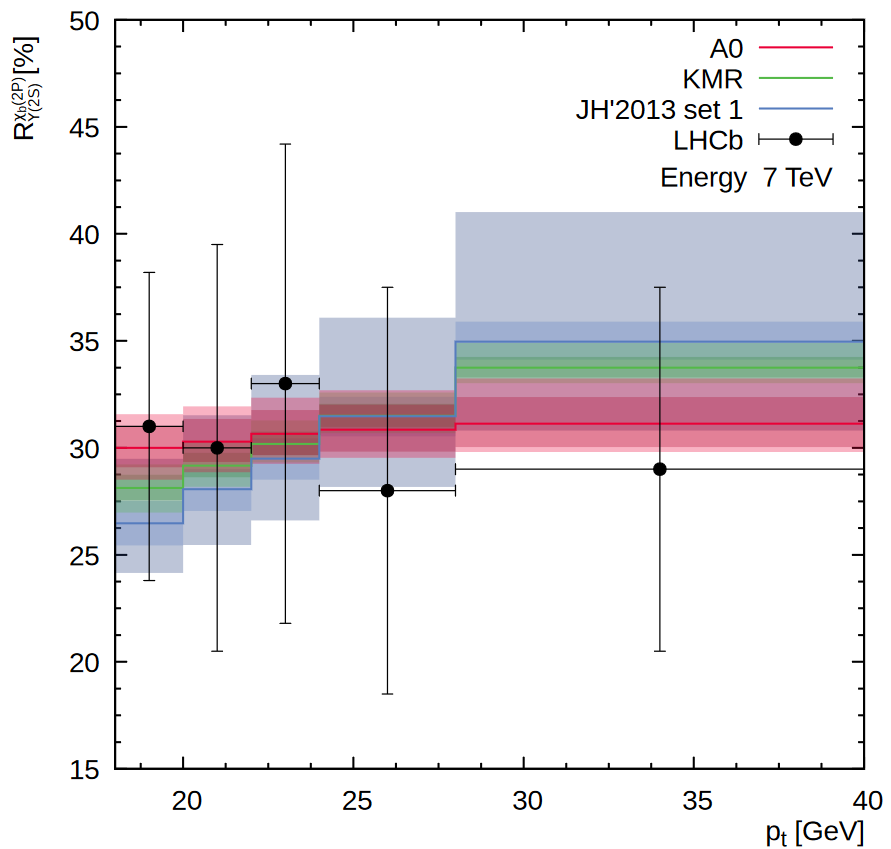}
\includegraphics[width=7.0cm]{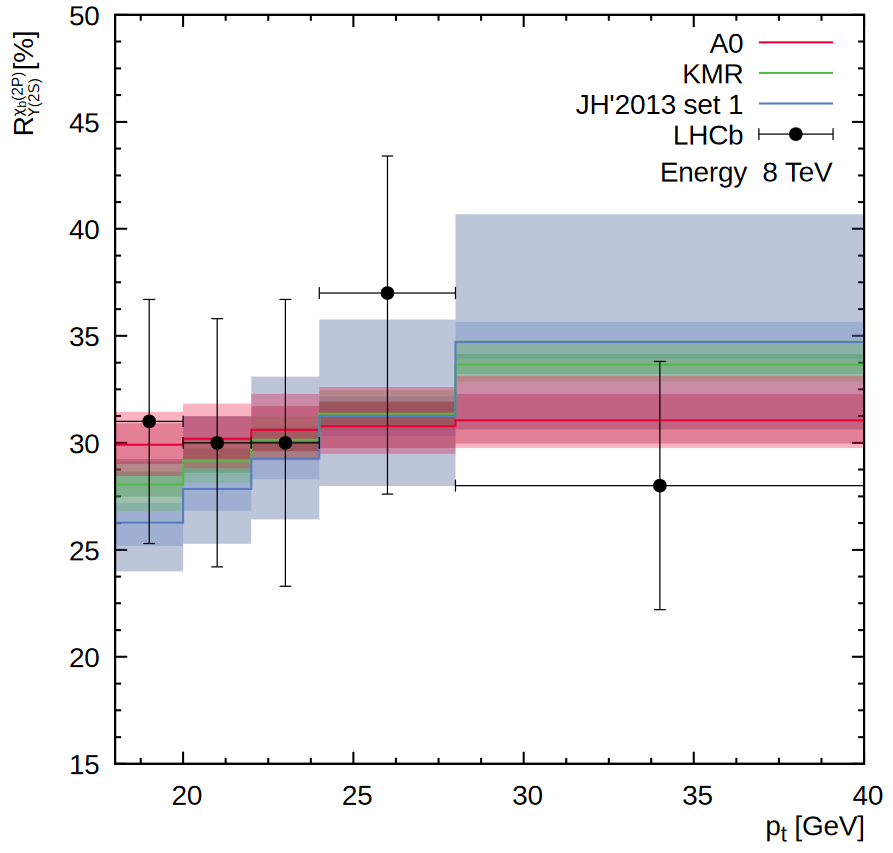}
\includegraphics[width=7.0cm]{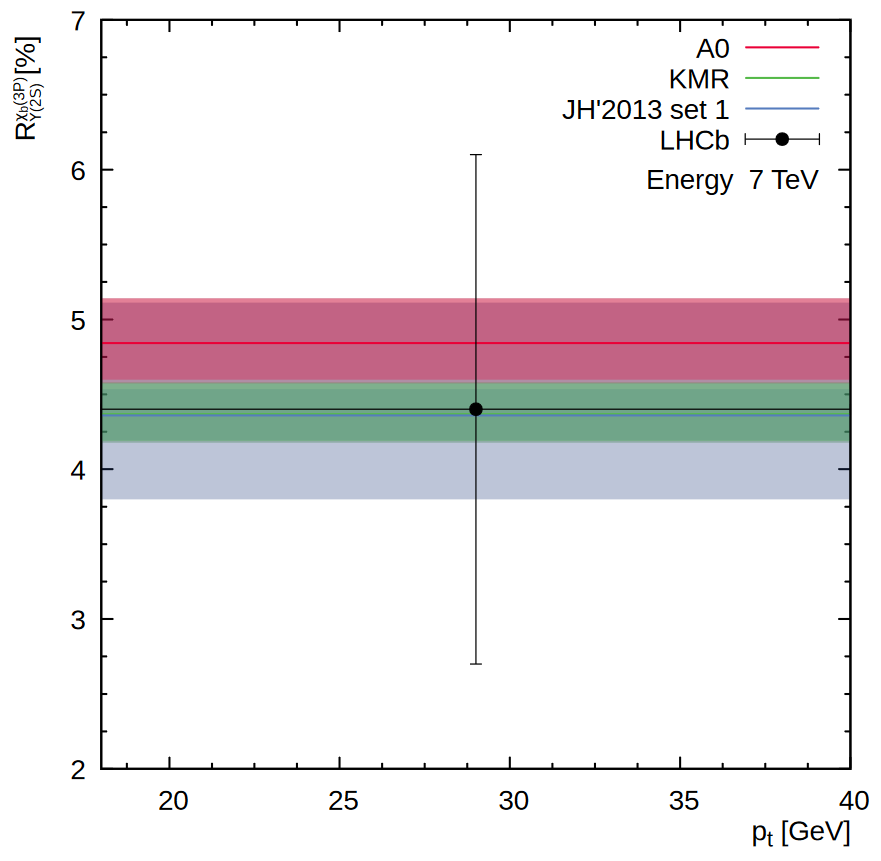}
\includegraphics[width=7.0cm]{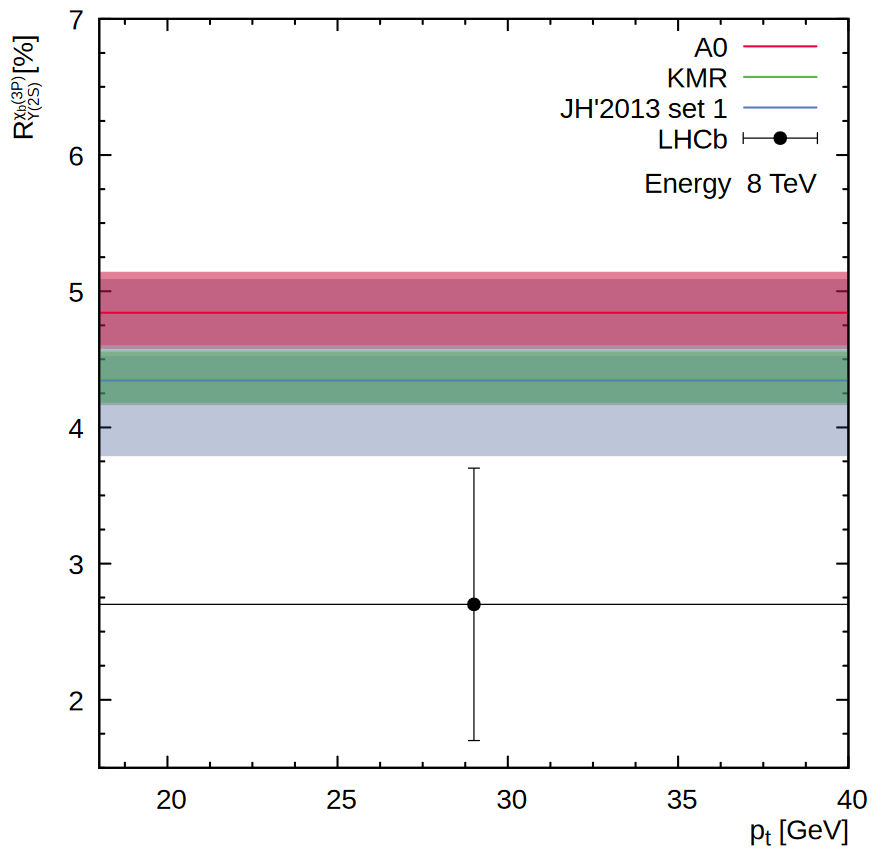}
\caption{The ratio $R^{\chi_b(mP)}_{\Upsilon(2S)}$
  calculated as function of $\Upsilon(2S)$ transverse momentum.
  Notation of all histograms is the same as in Fig.~\ref{fig1}.
  The experimental data are from LHCb \cite{lhcbr}.}
\label{fig3}
\end{center}
\end{figure}

All the data used in the fits are compared with our predictions 
in Figs.~\ref{fig1} --- \ref{fig3}.
The shaded areas represent the theoretical uncertainties of our 
calculations, 
which include the scale uncertainties, uncertainties coming from the NME 
fitting procedure and uncertainties connected with the 
choice of the intermediate color-octet mass, 
added in quadrature. 
To estimate the scale uncertainties
the standard variations 
$\mu_R\to 2\mu_R$ or $\mu_R\to\mu_R/2$ were introduced 
with replacing the A0 and JH'2013 set 1 gluon densities
by A0$+$ and JH'2013 set 1$+$, or 
by A0$-$ and JH'2013 set 1$-$ ones.
This was done to preserve the intrinsic correspondence between 
the TMD gluon set and the scale used in the CCFM evolution \cite{A0set,jhs}.
To estimate the uncertainties connected with the 
intermediate color-octet mass we have varied 
amount of energy $E$ emitted in the course of transition of 
unbound color octet $b\bar b$ pair into the observed 
bottomonium by a factor of $2$
around its default value $E = \Lambda_{\rm QCD}$.
One can see that we have achieved a reasonably good 
description of the CMS\cite{cms1,cms2} and ATLAS\cite{atlas} 
data in a whole $p_T$ range within the experimental and 
theoretical uncertainties for the $\Upsilon(2S)$ transverse 
momentum distributions. The ratios $R^{\chi_b(2P)}_{\Upsilon(2S)}$
and $R^{\chi_b(3P)}_{\Upsilon(2S)}$  measured by the LHCb Collaboration\cite{lhcbr} 
at $\sqrt s = 7$ and $8$~TeV are also reproduced well, see Fig.~4.

Finally, we have checked our results with the data, not included into the fit procedure: namely,
rather old CDF data \cite{cdf} taken at the $\sqrt s = 1.8$~TeV
and LHCb data \cite{lhcb1,lhcb2} taken in the forward rapidity region $2 < y < 4.5$
at $\sqrt s = 7$, $8$ and $13$~TeV (see Fig.~5). 
As one can see, we acceptably describe all the data above.
Moreover, we find that the KMR gluon is only one which is able to 
reproduce the measurements in the low $p_T$ region.

\subsection{$\Upsilon(2S)$ polarization} \indent

The polarization of any vector meson can be described 
with three parameters $\lambda_\theta$, $\lambda_\phi$ and 
$\lambda_{\theta\phi}$, which determine the spin density matrix of a 
meson decaying into a lepton pair and can be measured experimentally. 
The double differential angular distribution of the decay leptons 
can be written as \cite{siglam}:	
\begin{equation}
  {{d\sigma}\over{d\cos\theta^*d\phi^*}} \sim {{1}\over{3+\lambda_\theta}}(1 + \lambda_\theta\cos^2\theta^* + \lambda_\phi\sin^2\theta^*\cos2\phi^* + \lambda_{\theta\phi}\sin2\theta^*\cos\phi^*), 
  \label{eqlam}
\end{equation}

\noindent where $\theta^*$ and $\phi^*$ are the polar and azimuthal 
angles of the decay lepton measured in the meson rest frame.
The case of 
$(\lambda_\theta$, $\lambda_\phi$, $\lambda_{\theta \phi}) = (0,0,0)$
corresponds to unpolarized state, while 
$(\lambda_\theta$, $\lambda_\phi$, $\lambda_{\theta \phi}) = (1,0,0)$
and $(\lambda_\theta$, $\lambda_\phi$, $\lambda_{\theta \phi}) = (-1,0,0)$
refer to fully transverse and fully longitudinal polarizations. 

The CMS Collaboration has measured all of these polarization parameters for $\Upsilon(2S)$ mesons
as functions of their transverse momentum 
in three complementary frames: the Collins-Soper, helicity and 
perpendicular helicity ones at $\sqrt{s} = 7$ TeV \cite{cmslam}. 
The frame-independent parameter
$\tilde \lambda = (\lambda_\theta + 3\lambda_\phi)/(1 - \lambda_\phi)$ 
has been additionally studied.
The CDF Collaboration has measured $\lambda_\theta$ and $\tilde \lambda$ parameters  
in the helicity frame at $\sqrt{s} = 1.96$~TeV \cite{cdf2}.
As it was done previously \cite{upsilonI}, to estimate $\lambda_\theta$, $\lambda_\phi$, 
$\lambda_{\theta\phi}$ and $\tilde \lambda$ we generally follow the experimental procedure. 
We collect the simulated events in the kinematical region defined by
the experimental setup, generate the decay lepton angular
distributions according to the production and decay matrix elements 
and then apply a three-parametric fit based on~(12).

Our predictions are shown in Figs.~\ref{fig5} --- \ref{fig8}. 
The calculations were done using the A0 
gluon density which provides a best description of the measured 
$\Upsilon(2S)$ transverse momenta distributions.
As one can see, we find only weak or zero polarization in the all kinematic regions,
that agrees with the CMS and CDF measurements. 
The similar results we have obtained earlier for charmonia ($J/\psi$, $\psi^\prime$)
and $\Upsilon(3S)$ polarization \cite{baranovLip,baranovLip2,upsilonI}.
Thus, we conclude that the approach \cite{baranov}, which is a corner stone 
of our consideration, results in a self-consistent and simultaneous 
description of the entire charmonia family, $\Upsilon(2S)$ and $\Upsilon(3S)$ production data
and therefore can provide an easy and natural solution to the long-standing
quarkonia production and polarization puzzle.

\section{Conclusion} \indent

We have considered the $\Upsilon(2S)$ production at the Tevatron and LHC
in the framework of $k_T$-factorization 
approach. Our consideration was based on the off-shell 
production amplitudes for hard partonic subprocesses 
(including both color-singlet and color-octet contributions),  
NRQCD formalism for the formation of bound states
and TMD gluon densities in a proton. The latter were derived 
from the CCFM evolution equation and KMR scheme.
Treating the nonperturbative color octet transitions in terms of 
multipole radiation theory and
taking into account feed-down contributions from the 
radiative $\chi_b(3P)$ and $\chi_b(2P)$ decays and contribution from $\Upsilon(3S)$ decays,
we extracted
long-distance 
non-perturbative NRQCD matrix elements for
$\Upsilon(2S)$ and $\chi_b(2P)$ mesons 
from a fit to $\Upsilon(2S)$ transverse momentum distributions 
measured by the CMS and ATLAS Collaborations at  
$\sqrt s = 7$ and $13$~TeV
and from the relative production rates $R^{\chi_b(2P)}_{\Upsilon(2S)}$ 
measured by the LHCb Collaboration at $\sqrt s = 7$ and $8$~TeV.
Then we
estimated polarization parameters $\lambda_\theta$, $\lambda_\phi$,  
$\lambda_{\theta \phi}$ and frame-independent parameter $\tilde \lambda$
which determine the spin density matrix of $\Upsilon(2S)$ mesons.
We show that treating the soft gluon emission as a 
series of explicit color-electric 
dipole transitions within the NRQCD leads to unpolarized 
$\Upsilon(2S)$ production at moderate and large transverse 
momenta, that is in agreement with the LHC data.

\section*{Acknowledgements} \indent

The authors thank S.P.~Baranov and M.A.~Malyshev for their 
interest, useful discussions and important remarks.
N.A.A. is supported by the Foundation for the Advancement of 
Theoretical Physics and Mathematics ``Basis'' (grant No.18-1-5-33-1) and by RFBR, project number 19-32-90096.
A.V.L. is grateful the DESY Directorate for the support
in the framework of Cooperation Agreement between MSU and DESY 
on phenomenology of the LHC processes and TMD parton densities.

{\bibliography{biblio}}

\begin{thebibliography}{10}

\bibitem{kramer}
M.~Kr{\"a}mer{,} Prog. Part. Nucl. Phys. {\bf 47}{,}~14 (2001).

\bibitem{lansberg}
J.P. Lansberg{,} Int. J. Mod. Phys. A {\bf 21}{,}~3857 (2006).

\bibitem{bramilla}
N.~Brambilla et~al.{,} Eur. Phys. J. C {\bf 71}{,} 1534~(2011).

\bibitem{gongLi}
B.~Gong{,} X. Q. Li{,} J.-X. Wang{,} Phys. Lett. B {\bf 673}{,}~197 (2009).

\bibitem{maWang}
Y.-Q. Ma{,} K. Wang{,} K.-T. Chao{,} Phys. Rev. Lett. {\bf 106}{,}~042002
  (2011).

\bibitem{butenschonkniehl}
M.~Butensch{\"o}n{,} B.A. Kniehl{,} Phys. Rev. Lett. {\bf 108}{,}~172002
  (2012).

\bibitem{chaoMa}
K.-T. Chao{,} Y.-Q. Ma{,} H.-S. Shao{,} K. Wang{,} Y.-J. Zhang{,} Phys. Rev.
  Lett. {\bf 108}{,}~242004 (2012).

\bibitem{gongWan}
B.~Gong{,} L.-P. Wan{,} J.-X. Wang{,} H.-F. Zhang{,} Phys. Rev. Lett. {\bf
  110}{,}~042002 (2013).

\bibitem{maWang2}
Y.-Q. Ma{,} K. Wang{,} K.-T Chao{,} H.-F. Zhang{,} Phys. Rev. D {\bf
  83}{,}~111503 (2011).

\bibitem{likhoded}
A.K. Likhoded{,} A.V. Luchinsky{,} S.V. Poslavsky{,} Phys. Rev. D {\bf
  90}{,}~074021 (2014).

\bibitem{likhoded2}
A.K. Likhoded{,} A.V. Luchinsky{,} S.V. Poslavsky{,} Mod. Phys. Lett. A {\bf
  30}{,}~1550032 (2015).

\bibitem{zhangYu}
H.-F. Zhang{,} L. Yu{,} S.-X. Zhang{,} L. Jia{,} Phys. Rev. D {\bf
  93}{,}~054033 (2016).

\bibitem{hanMa}
H.~Han{,} Y.-Q. Ma{,} C. Meng{,} H.-S. Shao{,} K.-T. Chao{,} Phys. Rev. Lett.
  {\bf 114}{,}~092005 (2015).

\bibitem{zhangSun}
H.-F. Zhang{,} Z. Sun{,} W.-L. Sang{,} R. Li{,} Phys. Rev. Lett. {\bf
  114}{,}~092006 (2015).

\bibitem{butenschonHe}
M.~Butensch{\"o}n{,} Z.G. He{,} B.A. Kniehl{,} Phys. Rev. Lett. {\bf
  114}{,}~092004 (2015).

\bibitem{biswal}
S.S. Biswal{,} K. Sridhar{,} J. Phys. G: Nucl. Part. Phys. {\bf 39}{,}~015008
  (2012).

\bibitem{bodwinBraaten}
G.~Bodwin{,} E. Braaten{,} G. Lepage{,} Phys. Rev. D {\bf 51}{,}~1125 (1995).

\bibitem{choLeibovich}
P.~Cho{,} A.K. Leibovich{,} Phys. Rev. D {\bf 53}{,} 150 (1996){;} Phys. Rev. D
  {\bf 53}{,}~6203 (1996).

\bibitem{upsilonI}
N.A. Abdulov{,} A.V. Lipatov{,} Eur. Phys. J. C {\bf 79}{,}~830 (2019).

\bibitem{cmslam}
CMS Collaboration{,} Phys. Rev. Lett. {\bf 110}{,}~081802 (2013).

\bibitem{cms3}
CMS Collaboration{,} Phys. Lett. B {\bf 727}{,}~381 (2013).

\bibitem{lhcb3}
LHCb Collaboration{,} Eur. Phys. J. C {\bf 74}{,}~2872 (2014).

\bibitem{cdf1}
CDF Collaboration{,} Phys. Rev. Lett. {\bf 99}{,}~132001 (2007).

\bibitem{cdf2}
CDF Collaboration{,} Phys. Rev. Lett. {\bf 108}{,}~151802 (2012).

\bibitem{faccioli}
P.~Faccioli{,} V. Knuenz{,} C. Lourenco{,} J. Seixas{,} H.K. Woehri{,} Phys.
  Lett. B {\bf 736}{,}~98 (2014).

\bibitem{baranov}
S.P. Baranov{,} Phys. Rev. D {\bf 93}{,}~054037 (2016).

\bibitem{baranovLip}
S.P. Baranov{,} A.V. Lipatov{,} Eur. Phys. J. C {\bf 79}{,}~621 (2019).

\bibitem{baranovLip2}
S.P. Baranov{,} A.V. Lipatov{,} Phys. Rev. D {\bf 100}{,}~114021 (2019).

\bibitem{cms1}
CMS Collaboration{,} Phys. Lett. B {\bf 749}{,}~14 (2015).

\bibitem{cms2}
CMS Collaboration{,} Phys. Lett. B {\bf 780}{,}~251 (2018).

\bibitem{atlas}
ATLAS Collaboration{,} Phys. Rev. D {\bf 87}{,}~052004 (2013).

\bibitem{lhcb1}
LHCb Collaboration{,} JHEP {\bf 1511}{,}~103 (2015).

\bibitem{lhcb2}
LHCb Collaboration{,} JHEP {\bf 1807}{,}~134 (2018).

\bibitem{gongWang}
B.~Gong{,} J.-X. Wang{,} H.-F. Zhang{,} Phys. Rev. D {\bf 83}{,}~114021 (2011).

\bibitem{wangMa}
K.~Wang{,} Y.-Q. Ma{,} K.-T. Chao{,} Phys. Rev. D {\bf 85}{,}~114003 (2012).

\bibitem{gongWan2}
B.~Gong{,} L.-P. Wan{,} J.-X. Wang{,} H.-F. Zhang{,} Phys. Rev. Lett. {\bf
  112}{,}~032001 (2014).

\bibitem{fengGong}
Y.~Feng{,} B. Gong{,} L.-P. Wan{,} J.-X. Wang{,} H.-F. Zhang{,} Chin. Phys. C
  {\bf 39}{,}~123102 (2015).

\bibitem{hanMa2}
H.~Han{,} Y.-Q. Ma{,} C. Meng{,} H.-S. Shao{,} Y.-J. Zhang{,} K.-T. Chao{,}
  Phys. Rev. D {\bf 94}{,}~014028 (2016).

\bibitem{sumkt1}
L.V. Gribov{,} E.M. Levin{,} M.G. Ryskin{,} Phys. Rep. {\bf 100}{,} 1 (1983){;}
  \\ E.M. Levin{,} M.G. Ryskin{,} Yu.M. Shabelsky{,} A.G. Shuvaev{,} Sov. J.
  Nucl. Phys. {\bf 53}{,}~657 (1991).

\bibitem{sumkt2}
S.~Catani{,} M. Ciafaloni{,} F. Hautmann{,} Nucl. Phys. B {\bf 366}{,} 135
  (1991){;} \\ J.C. Collins{,} R.K. Ellis{,} Nucl. Phys. B {\bf 360}{,}~3
  (1991).

\bibitem{bfkl}
E.A. Kuraev{,} L.N. Lipatov{,} V.S. Fadin{,} Sov. Phys. JETP {\bf 44}{,} 443
  (1976); \\ E.A. Kuraev{,} L.N. Lipatov{,} V.S. Fadin{,} Sov. Phys. JETP {\bf
  45}{,} 199 (1977); \\ I.I. Balitsky{,} L.N. Lipatov{,} Sov. J. Nucl. Phys.
  {\bf 28}{,}~822 (1978).

\bibitem{ccfm}
M.~Ciafaloni{,} Nucl. Phys. B {\bf 296}{,} 49 (1988); \\ S. Catani{,} F.
  Fiorani{,} G. Marchesini{,} Phys. Lett. B {\bf 234}{,} 339 (1990); \\ S.
  Catani{,} F. Fiorani{,} G. Marchesini{,} Nucl. Phys. B {\bf 336}{,} 18
  (1990); \\ G. Marchesini{,} Nucl. Phys. B {\bf 445}{,}~49 (1995).

\bibitem{angeles}
R.~Angeles-Martinez et~al.{,} Acta Phys. Polon. B {\bf 46}{,} 2501~(2015).

\bibitem{lhcbr}
LHCb Collaboration{,} Eur. Phys. J. C {\bf 74}{,}~3092 (2014).

\bibitem{bycling}
E.~Bycling{,} K. Kajantie{,} Particle Kinematics{,}~John Wiley and Sons (1973).

\bibitem{colorcs}
C.-H. Chang{,} Nucl. Phys. B {\bf 172}{,} 425 (1980){;} \\ E.L. Berger{,} D.L.
  Jones{,} Phys. Rev. D 2{\bf 3}{,} 1521 (1981){;} \\ R. Baier{,} R.
  R{\"u}ckl{,} Phys. Lett. B {\bf 102}{,} 364 (1981){;} \\ S.S. Gershtein{,}
  A.K. Likhoded{,} S.R. Slabospitsky{,} Sov. J. Nucl. Phys. {\bf 34}{,}~128
  (1981).

\bibitem{batuninCho}
A.V. Batunin{,} S.R. Slabospitsky{,} Phys. Lett B {\bf 188}{,} 269 (1987){;} \\
  P. Cho{,} M. Wise{,} S. Trivedi{,} Phys. Rev. D {\bf 51}{,}~R2039 (1995).

\bibitem{A0set}
H.~Jung{,} arXiv:hep ph/0411287.

\bibitem{jhs}
F.~Hautmann{,} H. Jung{,} Nucl. Phys. B {\bf 883}{,}~1 (2014).

\bibitem{KMR}
M.A. Kimber{,} A.D. Martin{,} M.G. Ryskin{,} Phys. Rev. D {\bf 63}{,} 114027
  (2001){;} \\ A.D. Martin{,} M.G. Ryskin{,} G. Watt{,} Eur. Phys. J. C {\bf
  31}{,} 73 (2003){;} \\ A.D. Martin{,} M.G. Ryskin{,} G. Watt{,} Eur. Phys. J.
  C {\bf 66}{,}~163 (2010).

\bibitem{nnpdf}
NNPDF Collaboration{,} Eur. Phys. J. C {\bf 77}{,}~663 (2017).

\bibitem{pegasus}
S.P. Baranov{,} A.V. Lipatov{,} M.A.~Malyshev{,} arXiv:1912.04204~[hep ph].

\bibitem{pdg}
PDG Collaboration{,} Phys. Rev. D {\bf 98}{,}~030001 (2018).

\bibitem{eichten}
E.J. Eichten{,} C.~Quigg{,} arXiv:1904.11542~[hep ph].

\bibitem{gnuplot}
www.gnuplot.info.

\bibitem{cdf}
CDF Collaboration{,} Phys. Rev. Lett. {\bf 88}{,}~161802 (2002).

\bibitem{siglam}
M.~Beneke{,}~M. Kr{\"a}mer{,} and M.~V{\"a}nttinen{,} Phys. Rev. D {\bf
  57}{,}~4258 (1998).

\end{thebibliography}
\newpage
	
	\begin{table}[H] \footnotesize
	\centering
	\begin{tabular}{lcccc}
	\hline
	\hline
	\\
	 &  A0 & JH'2013 set 1 &  KMR & NLO NRQCD \cite{fengGong}\\
	\\ 
	\hline
	\\
	$\langle\mathcal{O}^{\Upsilon(2S)}[{}^{3}S_1^{(1)}]\rangle$/GeV$^{3}$ & $4.15$ & $4.15$ & $4.15$ & $4.63$ \\
    \\
	$\langle\mathcal{O}^{\Upsilon(2S)}[{}^{1}S_0^{(8)}]\rangle$/GeV$^3$ & $0.0$ & $0.0$ & $0.0$ & $0.0062 \pm 0.0198$ \\	
	\\
	$\langle\mathcal{O}^{\Upsilon(2S)}[{}^{3}S_1^{(8)}]\rangle$/GeV$^3$ & $0.016 \pm 0.004$ & $0.002 \pm 0.006$ & $0.0019 \pm 0.0006$ & $0.0222 \pm 0.0024$ \\	
	\\
	$\langle\mathcal{O}^{\Upsilon(2S)}[{}^{3}P_0^{(8)}]\rangle$/GeV$^{5}$ & $0.014 \pm 0.009$ & $0.19 \pm 0.05$ & $0.14 \pm 0.02$ & $-0.0013 \pm 0.0043$ \\	
	\\
	$\langle\mathcal{O}^{\chi_{b0}(2P)}[{}^{3}P_0^{(1)}]\rangle$/GeV$^{5}$ & $2.61$ & $2.61$ & $2.61$ & $2.37$ \\
    \\
	$\langle\mathcal{O}^{\chi_{b0}(2P)}[{}^{3}S_1^{(8)}]\rangle$/GeV$^{3}$ & $0.0181 \pm 0.0007$ & $0.0117 \pm 0.0007$ & $0.0074 \pm 0.0004$ & $0.0109 \pm 0.0014$ \\
    \\
	\hline
	\hline
	\end{tabular}
	\caption{The NMEs for $\Upsilon(2S)$ and $\chi_b(2P)$ mesons as determined from our 
	fit at $p_T^{\rm cut} = 10$~GeV. The NMEs obtained in the NLO NRQCD\cite{fengGong} are shown for comparison.}
	\label{tab1}
	\end{table}
	
 	\begin{table}[H] \footnotesize
	\centering
	\begin{tabular}{lcccc}
	\hline
	\hline
	\\
	7 TeV & $p_T^{\rm cut} = 10$~GeV & $p_T^{\rm cut} = 12$~GeV & $p_T^{\rm cut} = 15$~GeV & $p_T^{\rm cut} = 17$~GeV\\
	\\ 
	\hline
	\\
	A0 & $1.34$ & $1.29$ & $1.25$ & $1.31$ \\	
	\\
	JH'2013 set 1 & $2.84$ & $2.32$ & $1.94$ & $1.94$ \\	
	\\
	KMR & $1.62$ & $1.63$ & $1.67$ & $1.76$ \\	
	\\
	\hline
	\\
	7 + 13 TeV & $p_T^{\rm cut} = 10$~GeV & $p_T^{\rm cut} = 12$~GeV & $p_T^{\rm cut} = 15$~GeV & $p_T^{\rm cut} = 17$~GeV\\
	\\
	\hline
	\\
	A0 & $2.72$ & $2.73$ & $2.77$ & $2.86$ \\	
	\\
	JH'2013 set 1 & $6.28$ & $6.08$ & $5.99$ & $6.14$ \\	
	\\
	KMR & $3.25$ & $3.32$ & $3.4$ & $3.52$ \\	
	\\
	\hline
	\hline
	\end{tabular}
	\caption{The dependence of the $\chi^2/d.o.f.$ achieved in the 
	fit procedure on the choice of $p_T^{\rm cut}$ at only $\sqrt s = 7$ TeV and at $7$ and $13$~TeV combined.}
	\label{tab2}
	\end{table}
	
\newpage	

\begin{figure}
\begin{center}
\includegraphics[width=7.0cm]{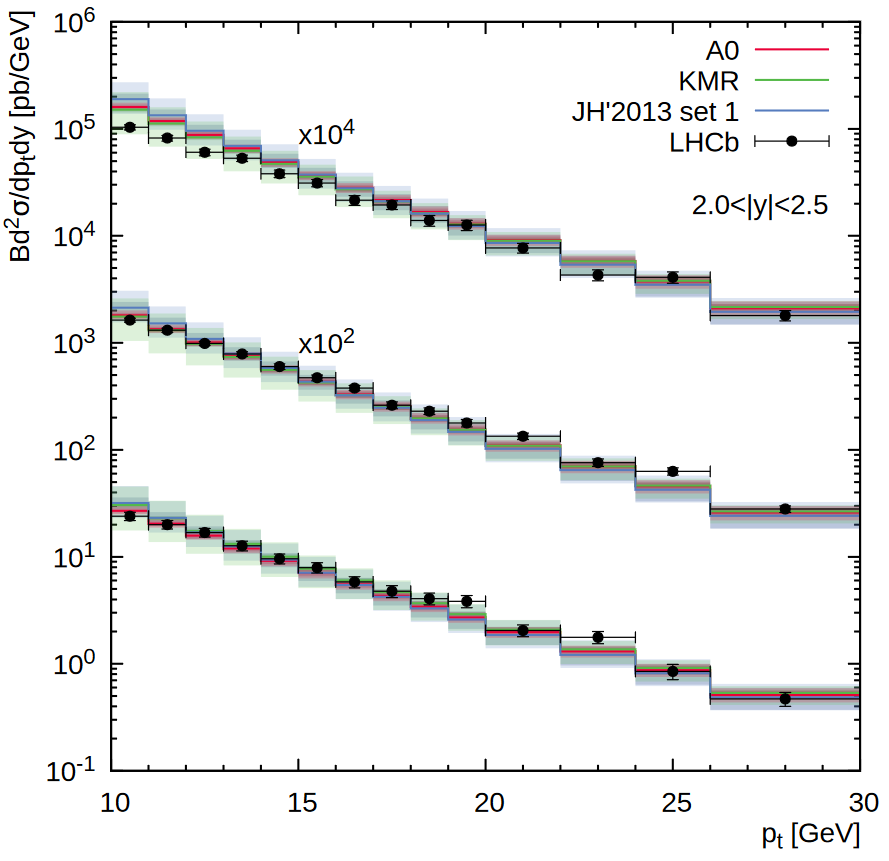}
\includegraphics[width=7.0cm]{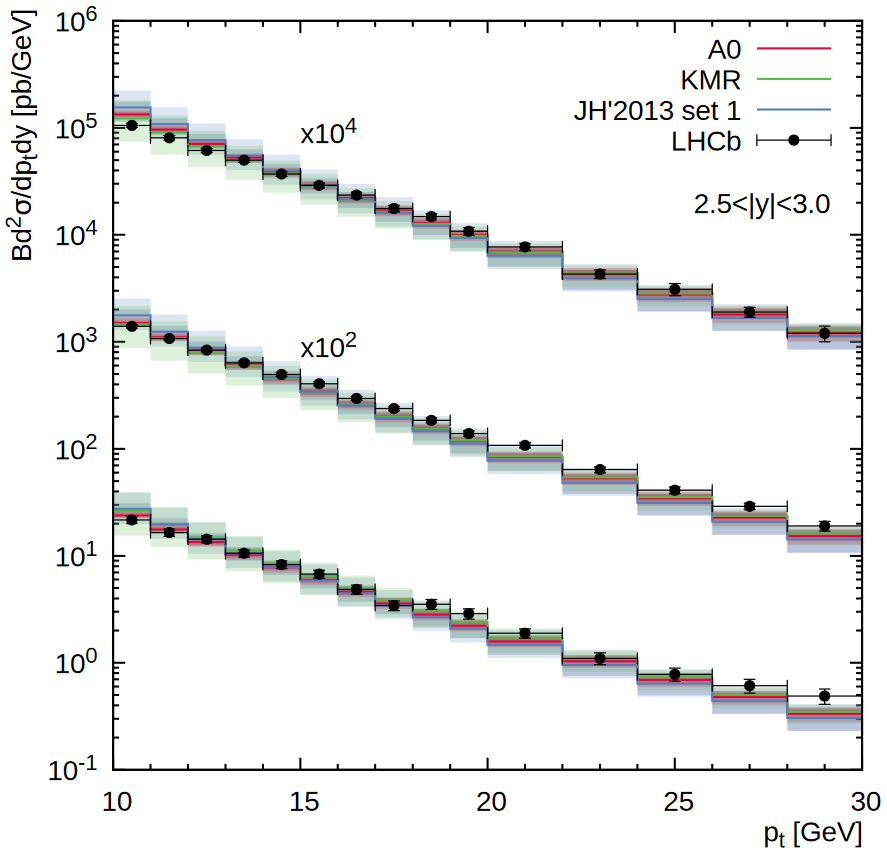}
\includegraphics[width=7.0cm]{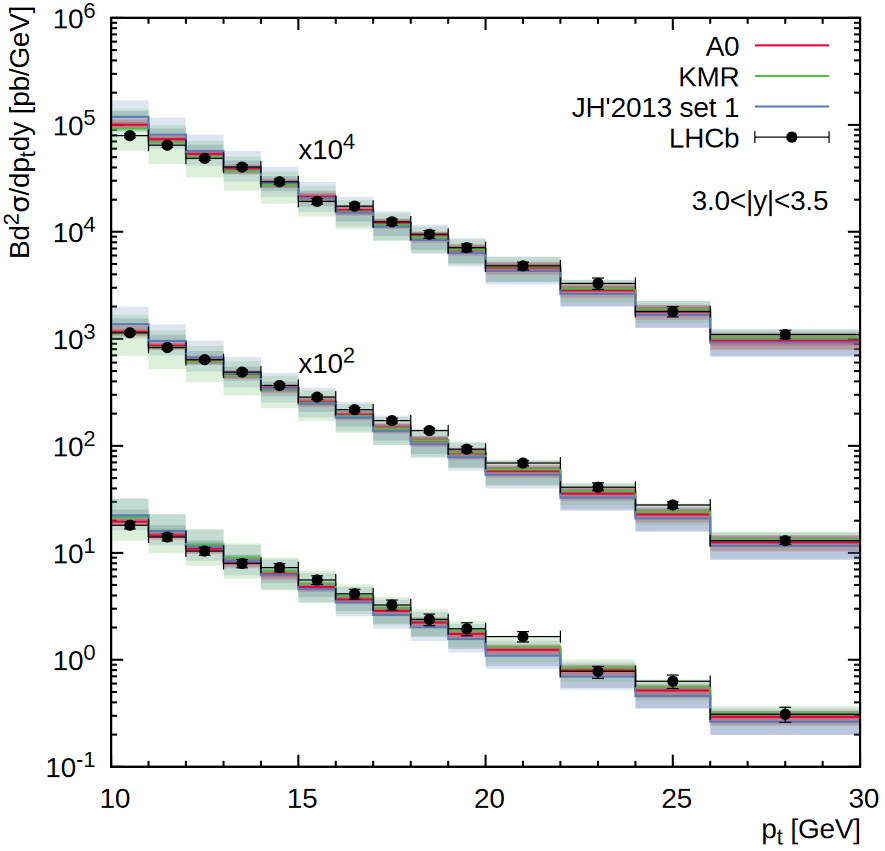}
\includegraphics[width=7.0cm]{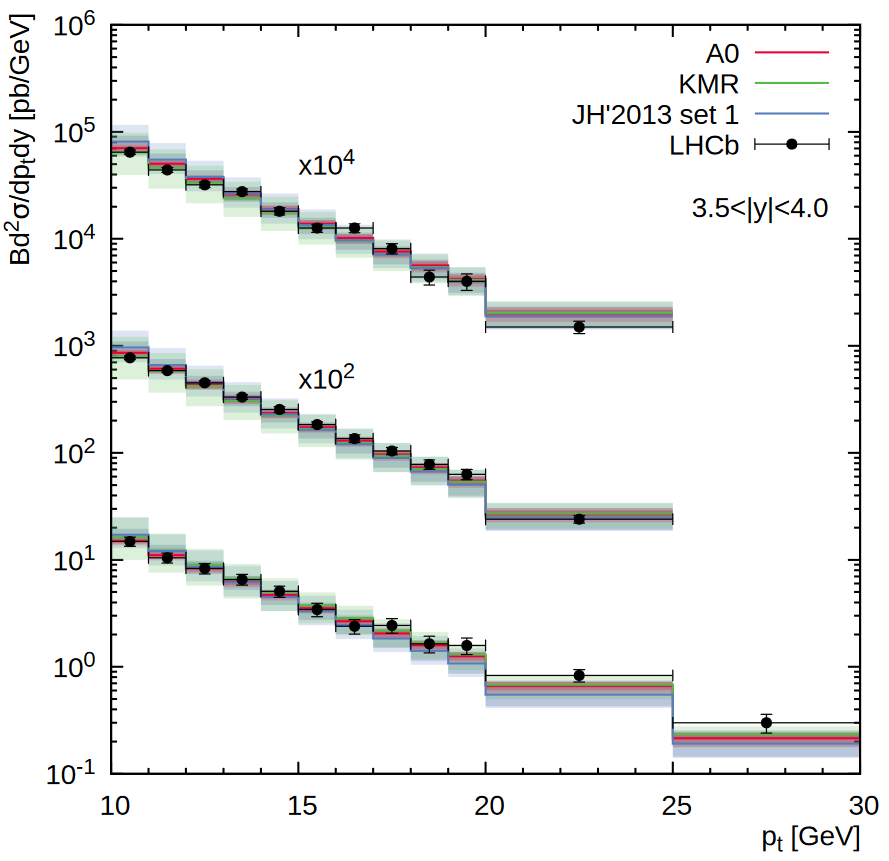}
\includegraphics[width=7.0cm]{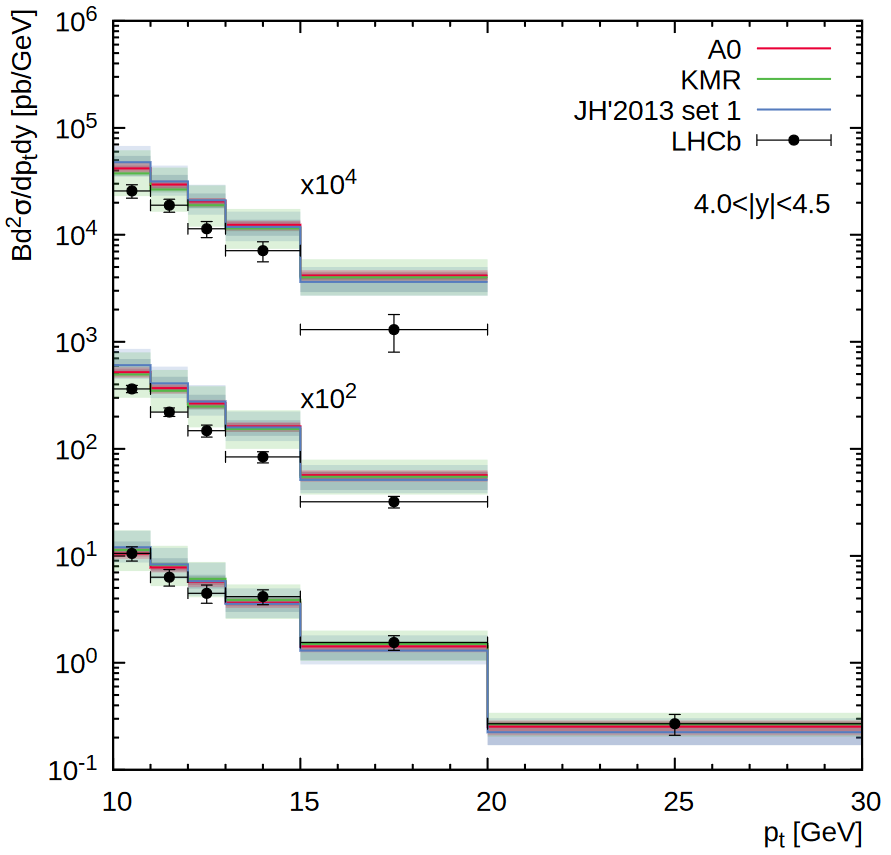}
\includegraphics[width=7.0cm]{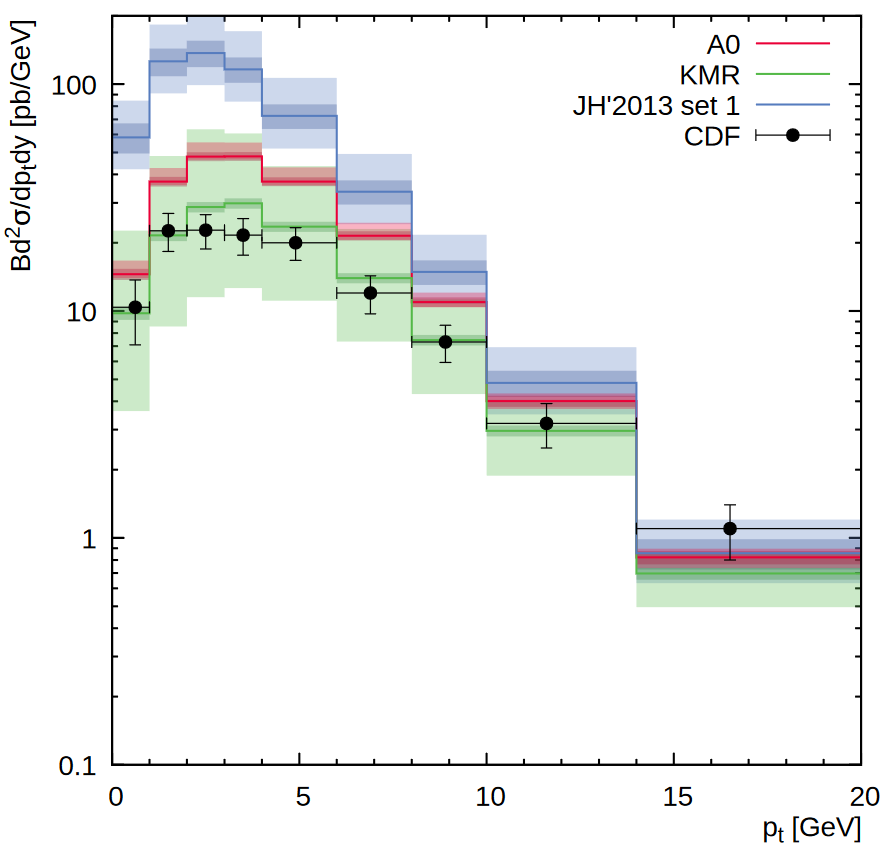}
\caption{Transverse momentum distribution of 
  inclusive $\Upsilon(2S)$ production calculated at $\sqrt s = 1.8$, 
  $7$, $8$ and $13$~TeV in the different rapidity regions. 
  Notation of all histograms is the same as in Fig.~\ref{fig1}.
  The experimental data are from CDF \cite{cdf} and LHCb \cite{lhcb1,lhcb2}.}
\label{fig4}
\end{center}
\end{figure}

\begin{figure}
\begin{center}
\includegraphics[width=7.0cm]{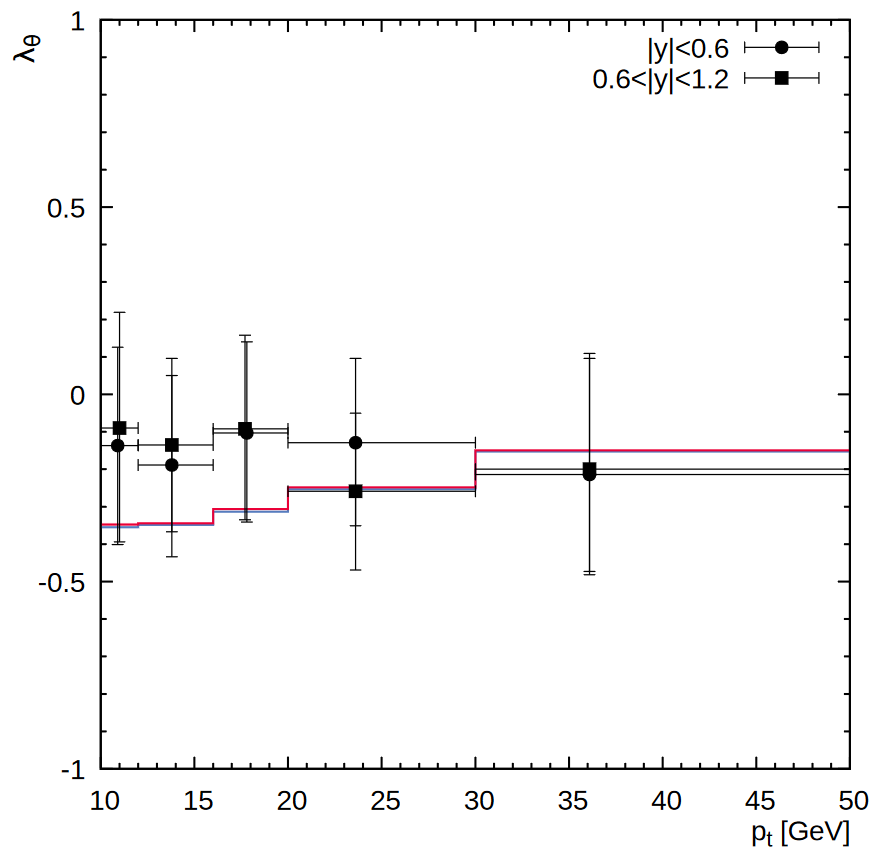}
\includegraphics[width=7.0cm]{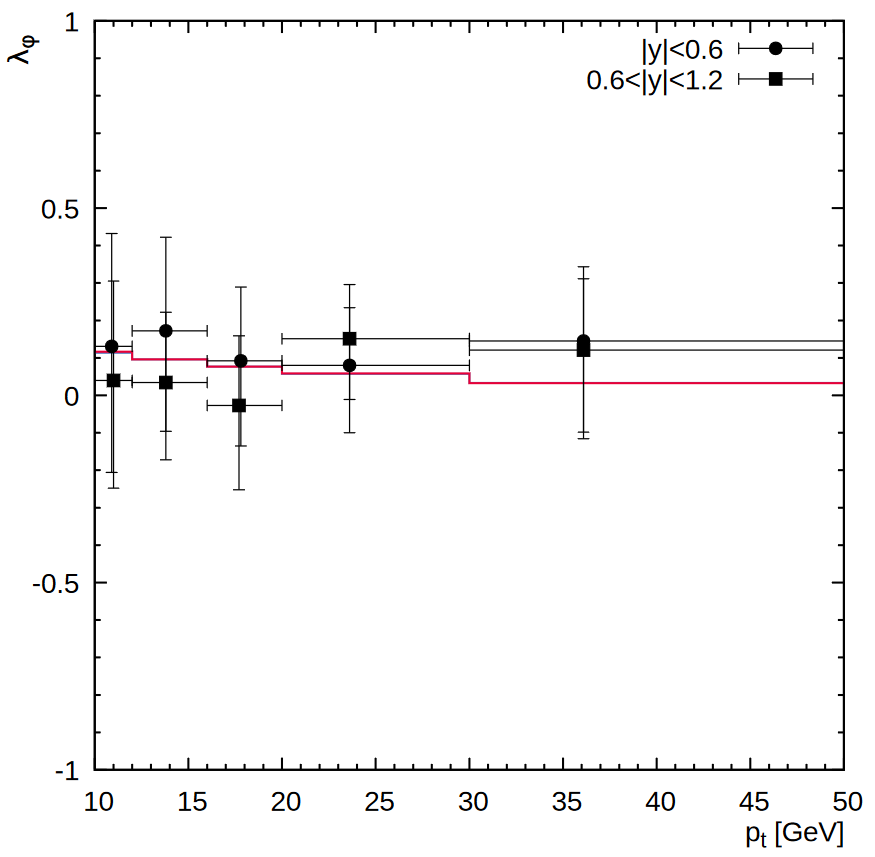}
\includegraphics[width=7.0cm]{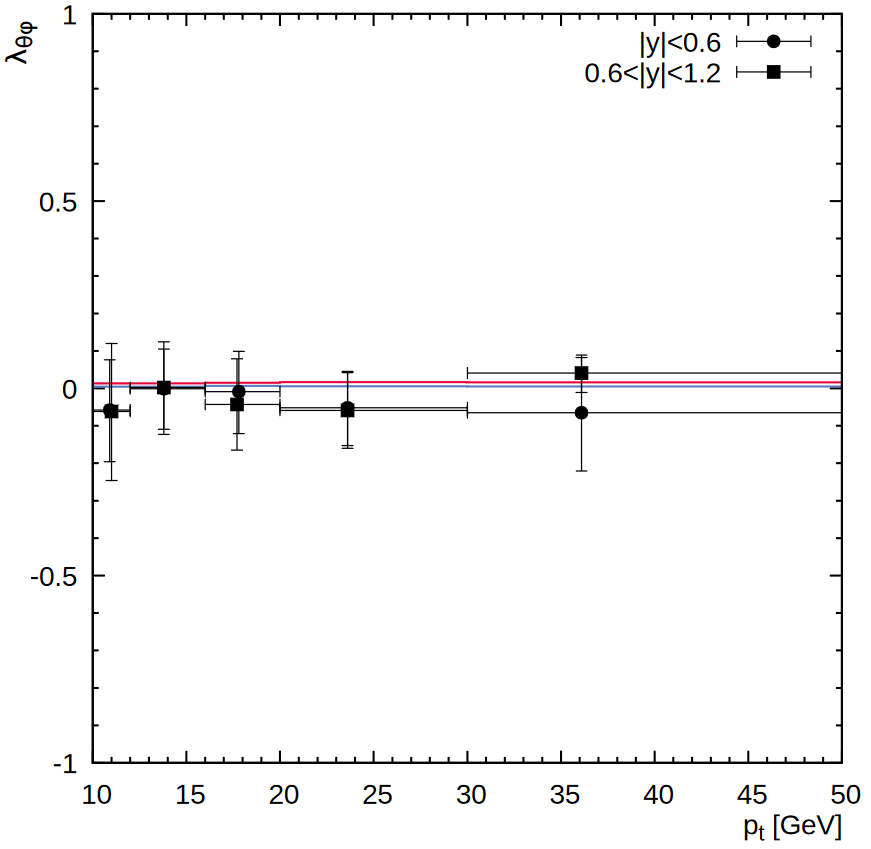}
\includegraphics[width=7.0cm]{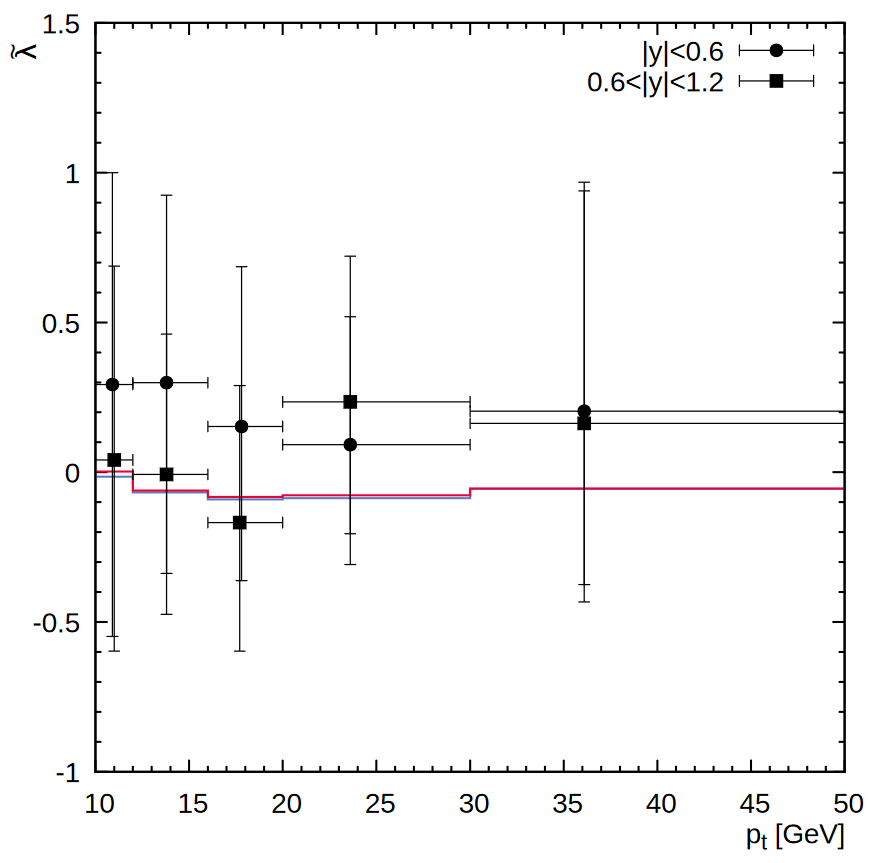}
\caption{The polarization parameters $\lambda_\theta$, 
  $\lambda_\phi$, $\lambda_{\theta\phi}$ and $\tilde\lambda$ of 
  $\Upsilon(2S)$ mesons calculated in the CS frame as function
  of its transverse momentum at $\sqrt{s} = 7$ TeV. 
  The A0 gluon density is used. The blue and red histograms 
  correspond to the predictions obtained at $|y|<0.6$ and 
  $0.6<|y|<1.2$, respectively. 
  The experimental data are from CMS\cite{cmslam}.}
\label{fig5}
\end{center}
\end{figure}

\begin{figure}
\begin{center}
\includegraphics[width=7.0cm]{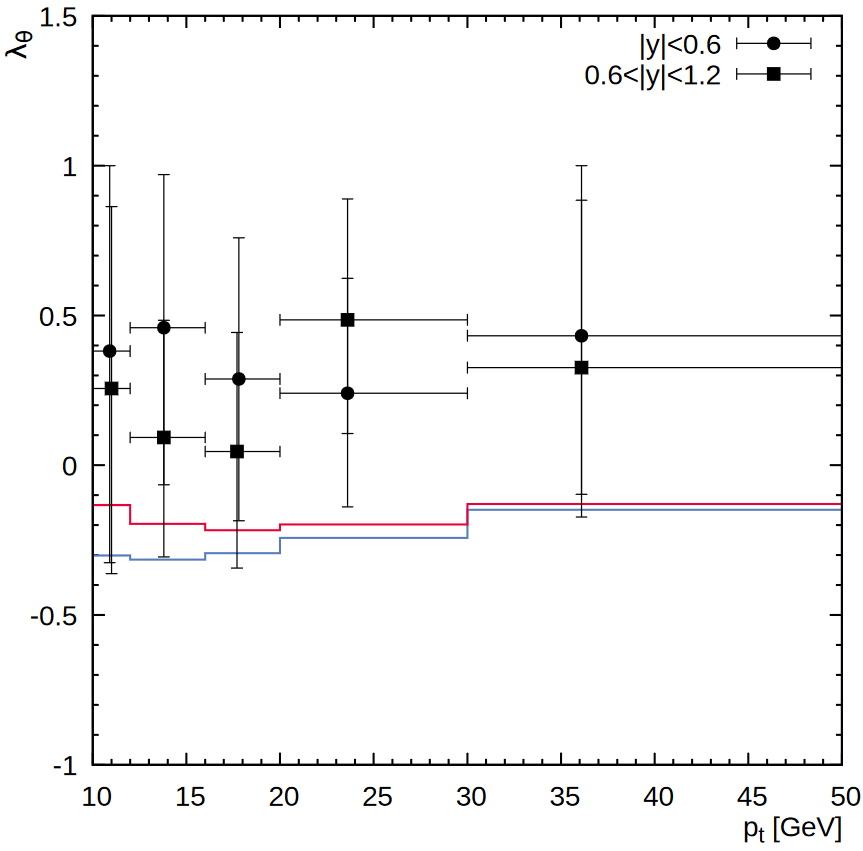}
\includegraphics[width=7.0cm]{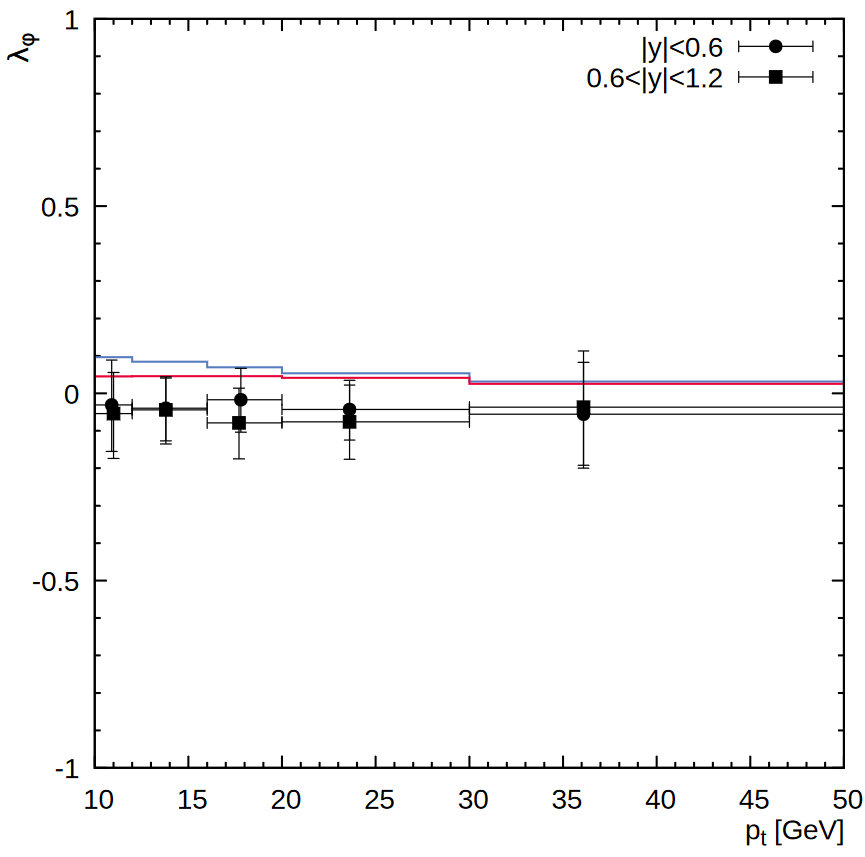}
\includegraphics[width=7.0cm]{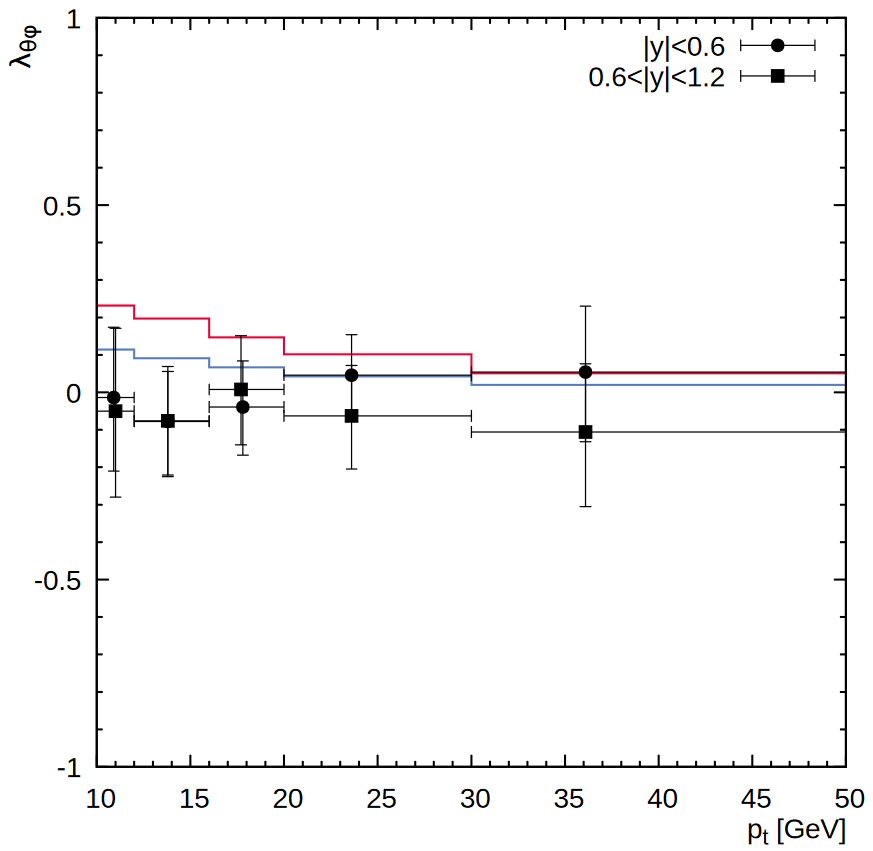}
\includegraphics[width=7.0cm]{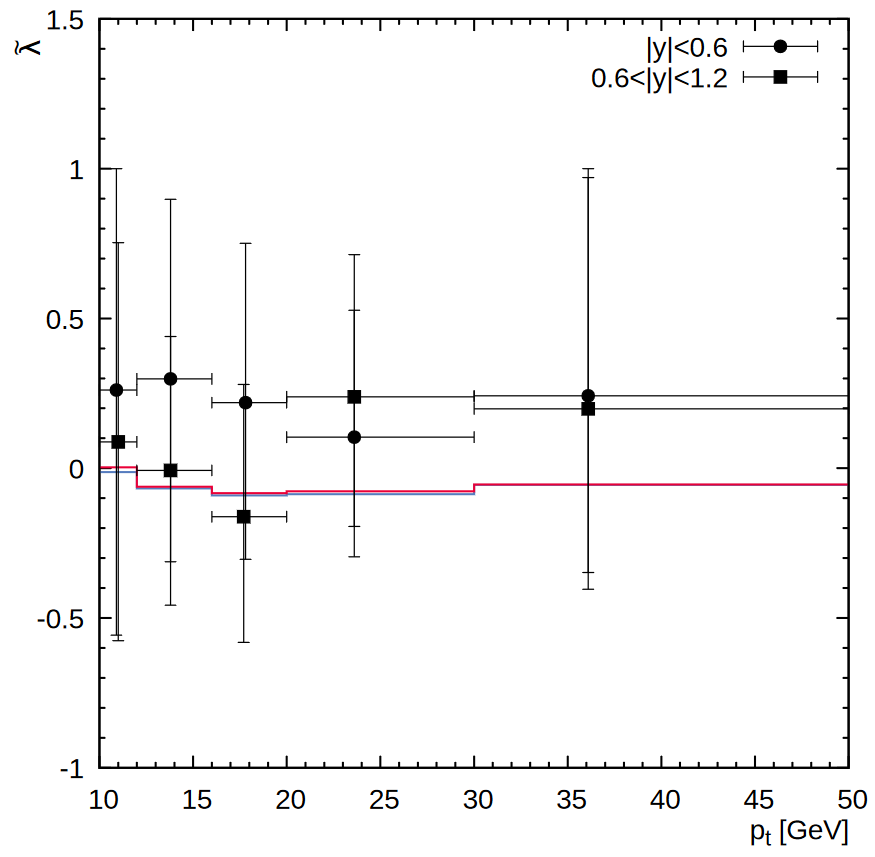}
\caption{The polarization parameters $\lambda_\theta$, 
  $\lambda_\phi$, $\lambda_{\theta\phi}$ and $\tilde\lambda$ of 
  $\Upsilon(2S)$ mesons calculated in the helicity frame as function
  of its transverse momentum at $\sqrt{s} = 7$ TeV. 
  Notation of all histograms is the same as in Fig.~\ref{fig5}.
  The experimental data are from CMS\cite{cmslam}.}
\label{fig6}
\end{center}
\end{figure}

\begin{figure}
\begin{center}
\includegraphics[width=7.0cm]{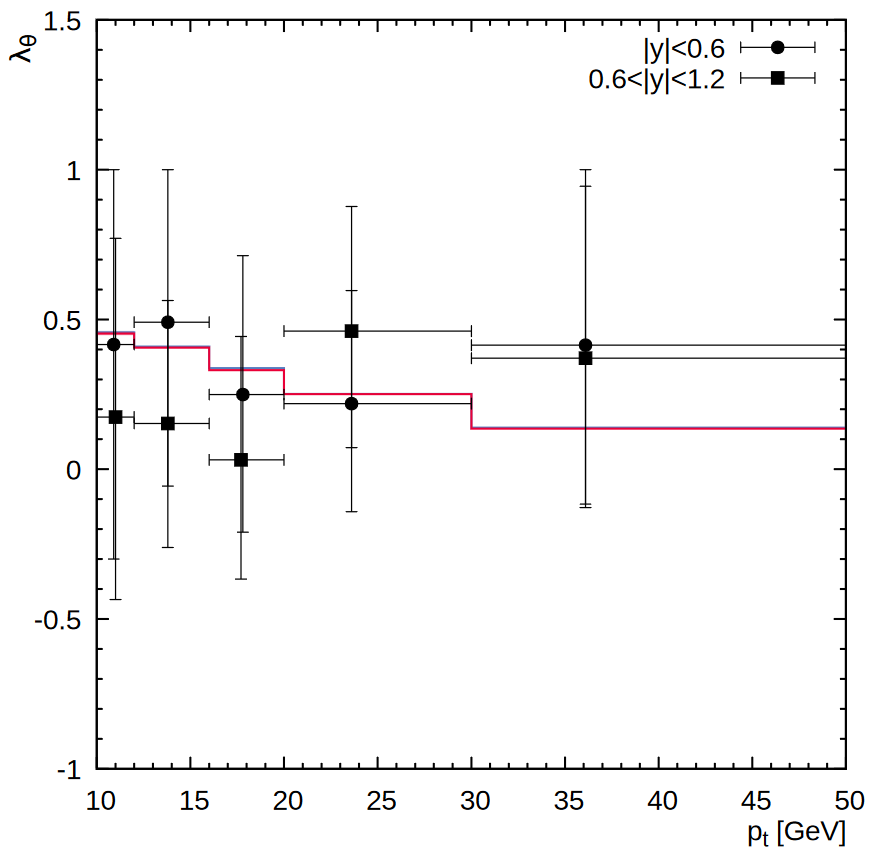}
\includegraphics[width=7.0cm]{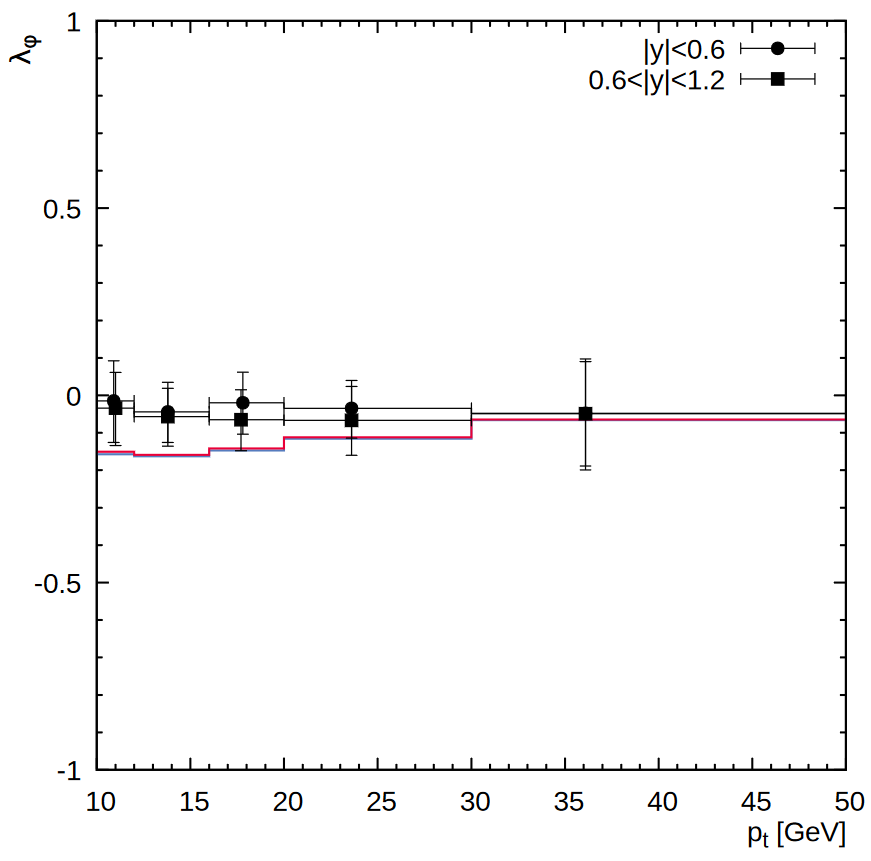}
\includegraphics[width=7.0cm]{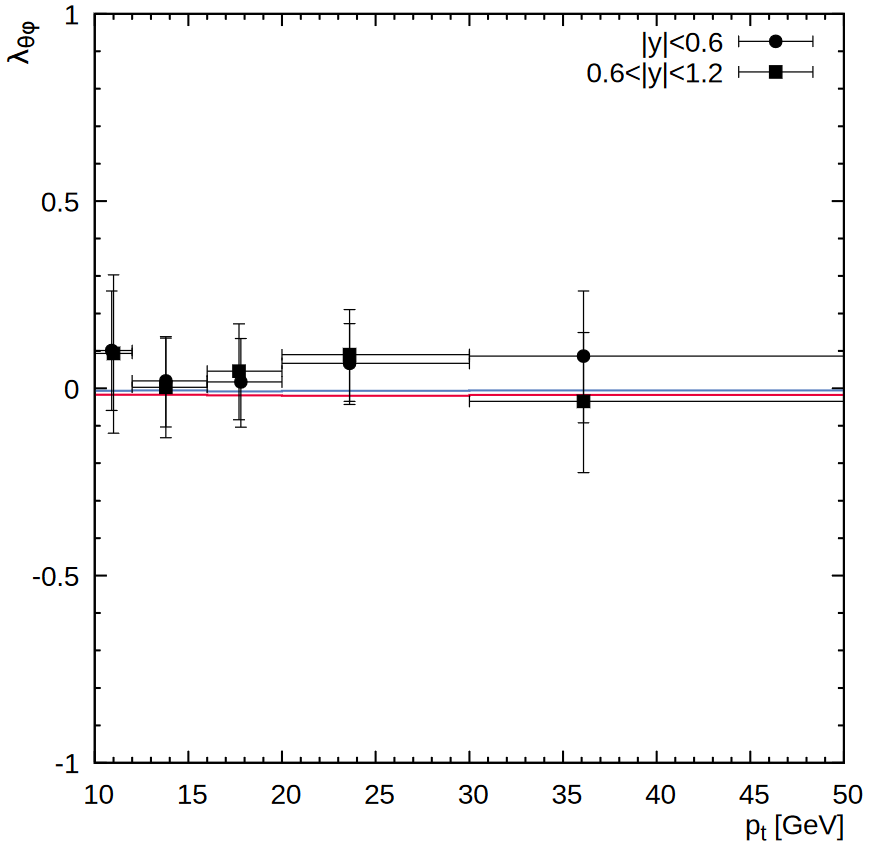}
\includegraphics[width=7.0cm]{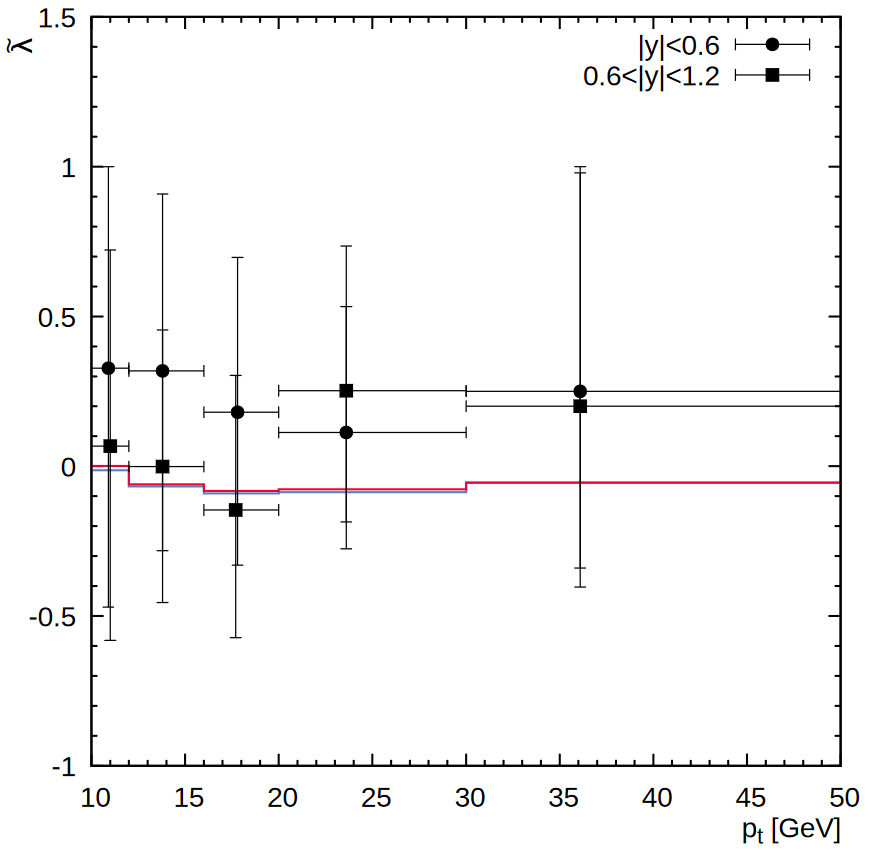}
\caption{The polarization parameters $\lambda_\theta$, 
  $\lambda_\phi$, $\lambda_{\theta\phi}$ and $\tilde\lambda$ of 
  $\Upsilon(2S)$ mesons calculated in the perpendicular 
  helicity frame as function
  of its transverse momentum at $\sqrt{s} = 7$ TeV. 
  Notation of all histograms is the same as in Fig.~\ref{fig5}.
  The experimental data are from CMS \cite{cmslam}.}
\label{fig7}
\end{center}
\end{figure}

\begin{figure}
\begin{center}
\includegraphics[width=7.0cm]{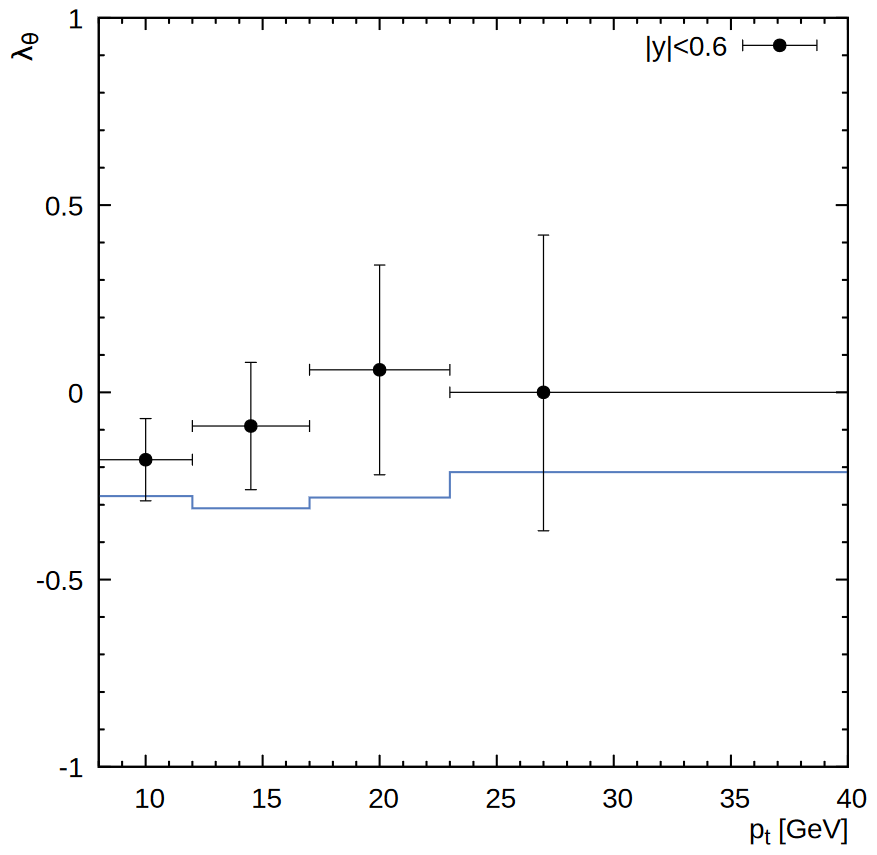}
\includegraphics[width=7.0cm]{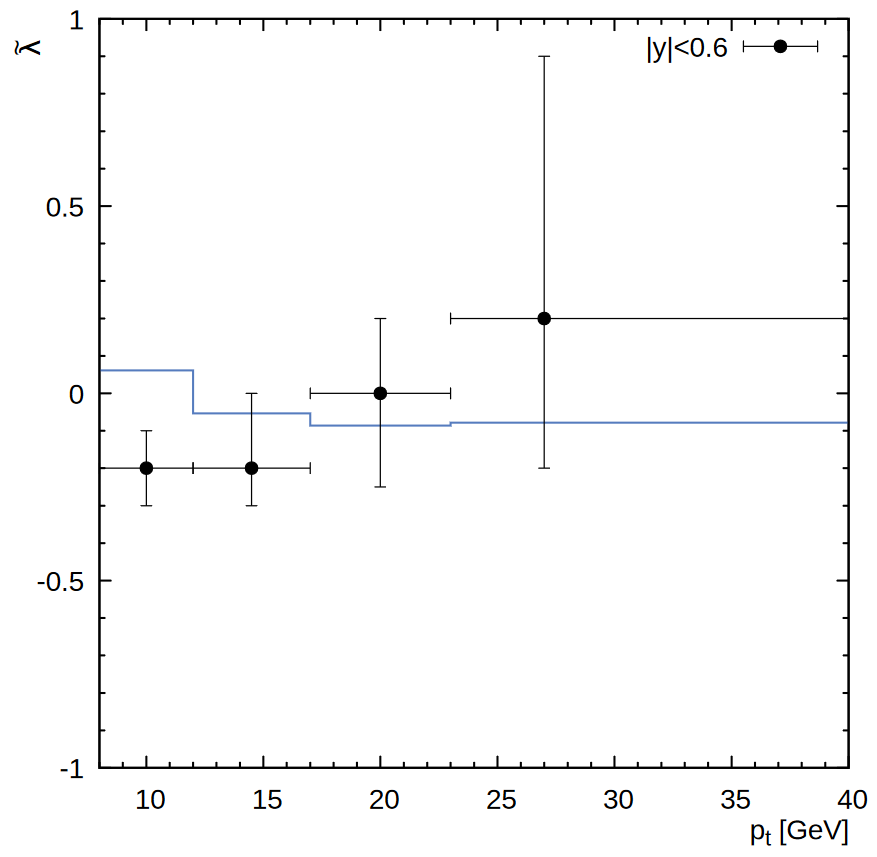}
\caption{The polarization parameters $\lambda_\theta$ and 
$\tilde\lambda$ of $\Upsilon(2S)$ mesons calculated in the 
helicity frame as function of its transverse momentum at 
$\sqrt{s} = 1.96$ TeV. Notation of all histograms is the same as in Fig.~\ref{fig5}.
  The experimental data are from CDF \cite{cdf2}.}
\label{fig8}
\end{center}
\end{figure}

\end{document}